\documentclass[%
 preprint,
 superscriptaddress,
 showkeys,
 amsmath,amssymb,
 aps,
]{revtex4-2}

\usepackage{graphicx}
\usepackage{dcolumn}
\usepackage{bm}
\usepackage{mathrsfs} 
\usepackage[colorlinks=true, allcolors=blue]{hyperref}
\newsavebox{\measurebox}
\usepackage{subcaption}
\usepackage[np]{numprint}
	\npthousandsep{,}
	\npdecimalsign{.}
\usepackage{multirow}

\begin{document}

\title{Reynolds Stress Anisotropy Tensor Predictions \\for Turbulent Channel Flow using Neural Networks}

\author{Jiayi Cai}
\affiliation{Service de Thermohydraulique et de M\'ecanique des Fluides, CEA, Universit\'e Paris-Saclay, Gif-sur-Yvette F-91191, France}
\affiliation{Laboratoire Interdisciplinaire des Sciences du Num\'erique LISN-CNRS, Orsay F-91403, France}
\author{Pierre-Emmanuel Angeli}
 \email{pierre-emmanuel.angeli@cea.fr}
\author{Jean-Marc Martinez}
\author{Guillaume Damblin}
\affiliation{Service de Thermohydraulique et de M\'ecanique des Fluides, CEA, Universit\'e Paris-Saclay, Gif-sur-Yvette F-91191, France}
\author{Didier Lucor}
\affiliation{Laboratoire Interdisciplinaire des Sciences du Num\'erique LISN-CNRS, Orsay F-91403, France}

\date{\today}

\begin{abstract}
The Reynolds-Averaged Navier-Stokes (RANS) approach remains a backbone for turbulence modeling due to its high cost-effectiveness. Its accuracy is largely based on a reliable Reynolds stress anisotropy tensor closure model. There has been an amount of work aiming at improving traditional closure models, while they are still not satisfactory to some complex flow configurations. In recent years, advances in computing power have opened up a new way to address this problem: the machine-learning-assisted turbulence modeling. In this paper, we employ neural networks to fully predict the Reynolds stress anisotropy tensor of turbulent channel flows at different friction Reynolds numbers, for both interpolation and extrapolation scenarios. Several generic neural networks of Multi-Layer Perceptron (MLP) type are trained with different input feature combinations to acquire a complete grasp of the role of each parameter. The best performance is yielded by the model with the dimensionless mean streamwise velocity gradient $\alpha$, the dimensionless wall distance $y^+$ and the friction Reynolds number $\mathrm{Re}_\tau$ as inputs. A deeper theoretical insight into the Tensor Basis Neural Network (TBNN) clarifies some remaining ambiguities found in the literature concerning its application of Pope's general eddy viscosity model. We emphasize the sensitivity of the TBNN on the constant tensor $\textbf{T}^{*(0)}$ upon the turbulent channel flow data set, and newly propose a generalized $\textbf{T}^{*(0)}$, which considerably enhances its performance. Through comparison between the MLP and the augmented TBNN model with both $\{\alpha, y^+, \mathrm{Re}_\tau\}$ as input set, it is concluded that the former outperforms the latter and provides excellent interpolation and extrapolation predictions of the Reynolds stress anisotropy tensor in the specific case of turbulent channel flow.
\end{abstract}


\maketitle

\section{Introduction}
Numerous researchers and engineers have been dedicated to understand the physics of turbulent fluid flows, through experiments or numerical resolution of the Navier-Stokes equations. Due to their high non-linearity, solving these equations is numerically difficult and expensive. The existence of turbulence makes it even more complex, because of its chaotic nature scaling widely in time and space. However, the understanding of turbulence is primordial, since not only does it commonly occur in every day life, for instance in waterfalls, fast flowing rivers or even blood flows to some extent, but it is also involved in numerous sophisticated technical designs and industrial processes, such as the nuclear power plant design, the aerodynamics of aircrafts, vehicles, etc. \cite{tennekes1972, pope2000, kronborg2022, international2022iaea, chapman1979}. 

From the numerical aspect, there exists namely three approaches for turbulence resolution, Direct Numerical Simulations (DNS), Large Eddy Simulations (LES) and Reynolds-Averaged Navier-Stokes (RANS) equations. The spatial and temporal resolution of DNS need to be high enough to capture the smallest eddies, so that no additional models need to be included in the solving process. The resolution of LES is about one order of magnitude coarser than DNS and some sub-grid models are applied to filter the eddies smaller than a certain size. These two methods are normally referenced as high fidelity simulations and require accurate numerical schemes with high computational cost. The RANS approach is on the other hand less accurate, since it only provides time-averaged solutions and includes turbulence closure models in all spatial scales. Therefore, no matter which model retained, there will exist epistemic errors with this approach (especially for configurations with separation effects and secondary flows \cite{johansson2002}), it is important to pursue more accurate closure modeling. Indeed, due to its high cost-effectiveness, the RANS method is still widely employed in engineering.

The modeling problem encountered in the RANS approach is known as the closure problem, which is caused by the appearance of the Reynolds stress tensor in the RANS equations due to the time-averaging operator and will be detailed hereinafter. As a result, closure models need to be included in order to solve for these unknowns. Traditionally, these models were developed by the combination of physical knowledge and experimental coefficient calibration on simple flow configurations. For example, the commonly used $k - \epsilon$ model includes five coefficients calibrated for plane jets and mixing layers, yet notably not suitable for axisymmetric jets \cite{launder1974}. 

In recent years, the development of computing resources has shed a new light on RANS modeling to overcome the deficiencies of traditional RANS closure models, via machine learning techniques \cite{duraisamy2019, brenner2019, brunton2020, duraisamy2021}. Among these machine-learning-assisted models, supervised algorithms dominate the published literature, which aim to learn the Reynolds stress tensor from high-fidelity data. A key attempt by Ling \textit{et al.} \cite{ling2016} has been to incorporate physical information into their proposed Tensor Basis Neural Network (TBNN). This novel architecture is physically based on Pope's generalized Reynolds stress model \cite{pope1975} on the purpose of preserving Galilean and rotational invariances. A series of studies have been carried out on the basis of this work. Fang \textit{et al.} \cite{fang2020} employed a more flexible neural network architecture, the Multi-Layer Perceptron (MLP), to account for additional physical properties of turbulent channel flow, which outperforms the TBNN; Sáez de Ocáriz Borde \textit{et al.} \cite{saezdeocarizborde2021} expanded this work upon the same flow configuration and designed a Convolutional Neural Network (CNN) architecture in order to better capture non-local turbulent effects; a recent work based on this architecture furtherly included multi-task learning and curriculum learning techniques to consider the turbulent duct flow \cite{saezdeocarizborde2022}. In addition to the above-mentioned trend of study in this field, several other directions exist, such as interpretability and generalizability analysis of the machine learning framework \cite{jiang2021, saezdeocarizborde2021}, uncertainty quantification \cite{emory2013, xiao2019}, conditioning problems \cite{wang2018, wu2019b} and full Partial Differential Equation (PDE) modeling \cite{jin2021, lucor2022}. It should be noted that the list provided here is by no means exhaustive given the amount of active studies in this area.  

The present work follows the research trend led by Ling \textit{et al.} \cite{ling2016} and aims to apply neural networks to fully predict the Reynolds stress tensor of turbulent channel flow, for both interpolation and extrapolation scenarios. In particular, we notice some ambiguities concerning the application of Pope's model for the TBNN that have not been clarified in the previous studies despite their promising success. For this sake, we focus on understanding and augmenting the TBNN architecture from a deeper physical insight and comparing its predictive capability with the generic neural network of MLP type upon the turbulent channel flow configuration. The paper is structured as follows. Section \ref{sec:background} provides theoretical background on turbulence RANS modeling and turbulent channel flow, followed by a review of deep learning techniques applied to this flow. After that, the methodology used for data preprocessing and neural network training is thoroughly explained in Section \ref{sec:method}. Results are then presented and discussed in Section \ref{sec:results}. Finally, conclusions and perspectives are given in Section \ref{sec:conclusions}.
 
\section{RANS modeling for turbulent channel flow} \label{sec:background}
The incompressible Navier-Stokes equations for unvarying viscosity can be stated as:
\begin{align} 
    \begin{cases}
\dfrac{\partial u_i}{\partial x_i}=0 \\[10pt]
\dfrac{\partial u_i}{\partial t}+{u}_j\dfrac{\partial {u}_i}{\partial x_j}=-\dfrac{1}{\rho}\dfrac{\partial {p}}{\partial x_i}+\nu\dfrac{\partial ^2 {u}_i}{\partial x_j \partial x_j} \label{eq:NS}
    \end{cases}   
\end{align}
\noindent where $u_i$, $x_i$, $t$, $\rho$, $p$ and $\nu$ are the components of the velocity vector, spatial coordinates, time, density, pressure and kinematic viscosity, respectively.

By decomposing the velocity and pressure into their mean and zero-mean fluctuating components ($u_i=\overline{u}_i+u^\prime_i$, $p=\overline{p}+p^\prime$) and then averaging the equations, we obtain the RANS equations:
\begin{align} 
    \begin{cases}
\dfrac{\partial {\overline{u}_i}}{\partial x_i}=0 \\[10pt]
\dfrac{\partial \overline{u}_i}{\partial t}+\overline{u}_j\dfrac{\partial \overline{u}_i}{\partial x_j}=-\dfrac{1}{\rho}\dfrac{\partial \overline{p}}{\partial x_i}+\nu\dfrac{\partial ^2 \overline{u}_i}{\partial x_j \partial x_j}-\dfrac{\partial \overline {u^\prime_iu^\prime_j}}{\partial x_j}. \label{eq:RANS}
    \end{cases}   
\end{align}

The major difficulty in RANS modeling is to solve for the unknown Reynolds stress tensor $\mathscr{R}_{ij}=\overline {u^\prime_iu^\prime_j}$ appearing in Eq.~\eqref{eq:RANS}. The Linear Eddy Viscosity Model (LEVM) is the most widely used model to tackle this closure problem, and can be approximated as:
\begin{equation}
\overline{u^\prime_iu^\prime_j}=-\nu_t(\dfrac{\partial \overline{u}_i}{\partial x_j}+\dfrac{\partial \overline{u}_j}{\partial x_i})+\dfrac{2}{3}k\delta_{ij}, \label{eq:BousR}
\end{equation}

\noindent where $\nu_t$ and $k$ are the eddy viscosity and turbulent kinetic energy, respectively; $\delta_{ij}$ denotes the Kronecker delta. This expression is known as the Boussinesq assumption \cite{boussinesq1897a}.

However, this simple linear relationship between the Reynolds stress tensor and the mean velocity gradients turns out to be inaccurate for some complex flows, especially involving secondary flows and curvature effects. An amount of Non-Linear Eddy Viscosity Models (NLEVM) at higher orders have been developed to capture these effects. For the sake of example, a Quadratic Eddy Viscosity Model (QEVM) simplified from Craft \textit{et al.} cubic model \cite{craft1996} can be written as:
\begin{equation}
\label{eq:craft}
\begin{split}
b_{ij}=&-\dfrac{\nu_t}{k}S_{ij}+C_1\dfrac{\nu_t}{\epsilon}\left(S_{ik}S_{kj}-\dfrac{1}{3}S_{kl}S_{kl}\delta_{ij}\right)+C_2\dfrac{\nu_t}{\epsilon}\left(R_{ik}S_{kj}+\dfrac{1}{3}R_{jk}S_{ki}\right) \\
&+C_3\dfrac{\nu_t}{\epsilon}\left(R_{ik}R_{jk}-\dfrac{1}{3}R_{kl}R_{kl}\delta_{ij}\right),
\end{split}
\end{equation}

\noindent where $\epsilon$ is the turbulent dissipation rate; $C_1$, $C_2$ and $C_3$ are parameters usually taken as constants, thanks to calibration on experiments or high-fidelity simulations upon simple flow configurations; $b_{ij}$, $S_{ij}$ and $R_{ij}$ are the Reynolds stress anisotropy, mean strain-rate and rotation-rate tensors, respectively:
\begin{equation}
b_{ij}=\dfrac{\overline{u^\prime_iu^\prime_j}}{2k}-\dfrac{1}{3}\delta_{ij}\label{eq:defb}
\end{equation} 
\begin{equation}
S_{ij}=\dfrac{1}{2}(\dfrac{\partial \overline{u}_i}{\partial x_j}+\dfrac{\partial \overline{u}_j}{\partial x_i})\label{eq:defS}
\end{equation}
\begin{equation}
R_{ij}=\dfrac{1}{2}(\dfrac{\partial \overline{u}_i}{\partial x_j}-\dfrac{\partial \overline{u}_j}{\partial x_i})\label{eq:defR}
\end{equation}

By combining Eqs.~\eqref{eq:BousR}, \eqref{eq:defb} and \eqref{eq:defS}, one can rewrite the LEVM in the RANS standard $k-\epsilon$ model \cite{launder1974} as follows:
\begin{equation}
\label{eq:BousB}
b_{ij}=-\dfrac{\nu_t}{k}S_{ij}=-C_\mu\dfrac{k}{\epsilon}S_{ij}
\end{equation}
\noindent which is equivalent to Craft \textit{et al.}'s cubic model shown in Eq.~\eqref{eq:craft} in first-order approximation. Here, $C_\mu$ is a calibrated parameter, generally taken as 0.09.

\subsection{General eddy viscosity model} \label{subsec:EVM}
One of the most generalized NLEVM was proposed by Pope \cite{pope1975} in order to extent the universality of RANS closure models. Pope's approach focused on the Reynolds stress anisotropy tensor $\textbf{b}$ and postulated that it can be expressed only in function of normalized tensors $\textbf{S}^*$ and $\textbf{R}^*$ for an homogeneous flow:
\begin{equation}
\textbf{b}=\textbf{b}(\textbf{S}^*, \textbf{R}^*), \label{eq:hypoPope}
\end{equation}

\noindent where $\textbf{S}^*$ and $\textbf{R}^*$ are respectively the mean strain-rate and the rotation-rate tensors normalized by a turbulent time scale formed with the turbulent kinetic energy and dissipation rate:
\begin{equation}
S_{ij}^*=\dfrac{1}{2}\dfrac{k}{\epsilon}(\dfrac{\partial \overline{u}_i}{\partial x_j}+\dfrac{\partial \overline{u}_j}{\partial x_i})\label{eq:defS*}
\end{equation}
\begin{equation}
R_{ij}^*=\dfrac{1}{2}\dfrac{k}{\epsilon}(\dfrac{\partial \overline{u}_i}{\partial x_j}-\dfrac{\partial \overline{u}_j}{\partial x_i})\label{eq:defR*}
\end{equation}

By furtherly supposing the function shown in Eq.~\eqref{eq:hypoPope} as a polynomial function and applying the Cayley-Hamilton theorem, Pope obtained the following closure model of $\textbf{b}$, which expresses it as a series of $n$ finite tensor polynomials:
\begin{equation}
\textbf{b}(\textbf{S}^*, \textbf{R}^*)=\sum_{n}g^{(n)}\left(\lambda_1^*, \lambda_2^*...\right)\textbf{T}^{*(n)} \label{generalPope}
\end{equation}
\noindent where $g^{(n)}$ are coefficient functions depending on physical independent invariants $\lambda_i^*$ and $\textbf{T}^{*(n)}$ are basis tensors depending on $\textbf{S}^*$ and $\textbf{R}^*$.

In the general case, there are five invariants and ten tensors $(1\leq n \leq10)$:
\begin{equation}
\lambda_1^*=\text{tr}(\textbf{S}^{*2})\, ,\quad \lambda_2^*=\text{tr}(\textbf{R}^{*2})\, ,\quad \lambda_3^*=\text{tr}(\textbf{S}^{*3})\, ,\quad \lambda_4^*=\text{tr}(\textbf{R}^{*2}\textbf{S})\, ,\quad \lambda_5^*=\text{tr}(\textbf{R}^{*2}\textbf{S}^{*2})\label{eq:PopeInvariant3D}
\end{equation}
\begin{align} 
    \begin{cases}
     	\textbf{T}^{*(1)} =  \textbf{S}^{*} &  \textbf{T}^{*(2)} =  \textbf{S}^{*}\textbf{R}^{*}-\textbf{R}^{*}\textbf{S}^{*}  \\
      	\textbf{T}^{*(3)} =  \textbf{S}^{*2}-\dfrac{\lambda_1^*}{3}\textbf{I}_3 & \textbf{T}^{*(4)} =  \textbf{R}^{*2}-\dfrac{\lambda_2^*}{3}\textbf{I}_3 \\
      	\textbf{T}^{*(5)} =  \textbf{R}^{*}\textbf{S}^{*2}-\textbf{S}^{*2}\textbf{R}^{*} & \textbf{T}^{*(6)} =  \textbf{R}^{*2}\textbf{S}^{*}+\textbf{S}^{*}\textbf{R}^{*2}-\dfrac{2\lambda_4^*}{3}\textbf{I}_3\\
      	\textbf{T}^{*(7)}=\textbf{R}^{*}\textbf{S}^{*}\textbf{R}^{*2}-\textbf{R}^{*2}\textbf{S}^{*}\textbf{R}^{*} & \textbf{T}^{*(8)}=\textbf{S}^{*}\textbf{R}^{*}\textbf{S}^{*2}-\textbf{S}^{*2}\textbf{R}^{*}\textbf{S}^{*} \\
      	\textbf{T}^{*(9)}=\textbf{R}^{*2}\textbf{S}^{*2}+\textbf{S}^{*2}\textbf{R}^{*2}-\dfrac{2\lambda_5^*}{3}\textbf{I}_3 & \textbf{T}^{*(10)}=\textbf{R}\textbf{S}^{*2}\textbf{R}^{*2}-\textbf{R}^{*2}\textbf{S}^{*2}\textbf{R}\\
     \end{cases}\label{eq:PopeTensor3D}
\end{align}
\noindent where $\textbf{I}_3$ denotes the identity tensor. 

It can be noticed that Pope's model is a generalized form of LEVM shown in Eq.~\eqref{eq:BousB} in first-order approximation and the QEVM shown in Eq.~\eqref{eq:craft} in second-order approximation. In particular, given Eq.~\eqref{eq:BousB}, the coefficient function $g^{(1)}$ is identified with $-C_\mu$ and should therefore be negative.

Specifically for flows where the mean velocity and the variation of mean quantities in one direction are zero, Pope demonstrates that only two invariants and a basis of three tensors are sufficient $(0\leq n \leq2)$:
\begin{equation}
\textbf{b}=g^{(0)}(\lambda_1^*, \lambda_2^*)\textbf{T}^{*(0)} + g^{(1)}(\lambda_1^*, \lambda_2^*)\textbf{T}^{*(1)} + g^{(2)}(\lambda_1^*, \lambda_2^*)\textbf{T}^{*(2)} \label{eq:2D_Pope}
\end{equation}

\noindent with
\begin{equation}
\lambda_1^*=\text{tr}(\textbf{S}^{*2})\, ,\quad \lambda_2^*=\text{tr}(\textbf{R}^{*2})\label{eq:PopeInvariant2D}
\end{equation}
\begin{equation}
    \begin{cases}

     	\textbf{T}^{*(0)} =  \dfrac{1}{2}\textbf{I}_2 - \dfrac{1}{3}\textbf{I}_3\\
     	\textbf{T}^{*(1)} =  \textbf{S}^{*} \\
      	\textbf{T}^{*(2)} =  \textbf{S}^{*}\textbf{R}^{*}-\textbf{R}^{*}\textbf{S}^{*}\\
    \end{cases}\label{eq:PopeTensor2D}
\end{equation}
\noindent where $\textbf{I}_2=\mathrm{diag}(1,1,0)$ or its permutations depending on the characterising direction of the flow. If, for example, there is zero mean velocity and invariance along the $x_3$ direction, then $\textbf{I}_2=\mathrm{diag}(1,1,0)$. One might notice that the invariants and basis tensors here are the same as those of the general model shown in Eqs.~\eqref{eq:PopeInvariant3D} and \eqref{eq:PopeTensor3D},  except for the choice of $\textbf{T}^{*(0)}$ instead of $\textbf{T}^{*(3)}$. In fact, it can be easily demonstrated that $\textbf{T}^{*(3)}=-\lambda_1^*\textbf{T}^{*(0)}$ under the restricted condition of this simplified case.

As Navier-Stokes equations, RANS equations and any other closure models, Pope's model, given in Eq.~\eqref{generalPope}, satisfies the Galilean and rotational invariances, which means that this model remains identical while undergoing a rectilinear and uniform motion or a rotation at a constant angle. The consideration of these invariances is fundamental since they are properties to which the fluid flow physically obeys. However, it should be kept in mind that Pope's model remains questionable in inhomogeneous region \cite{pope2000}, for example the near-wall region, where other parameters than $\textbf{S}^*$ and $\textbf{R}^*$ are needed to represent $\textbf{b}$. This issue sometimes referred to as a multi-value problem \cite{liu2021a} will be addressed hereinafter.

\subsection{Turbulent channel flow}  \label{subsec:TCF}
\begin{figure}[h]
\centering
\includegraphics[scale=0.3]{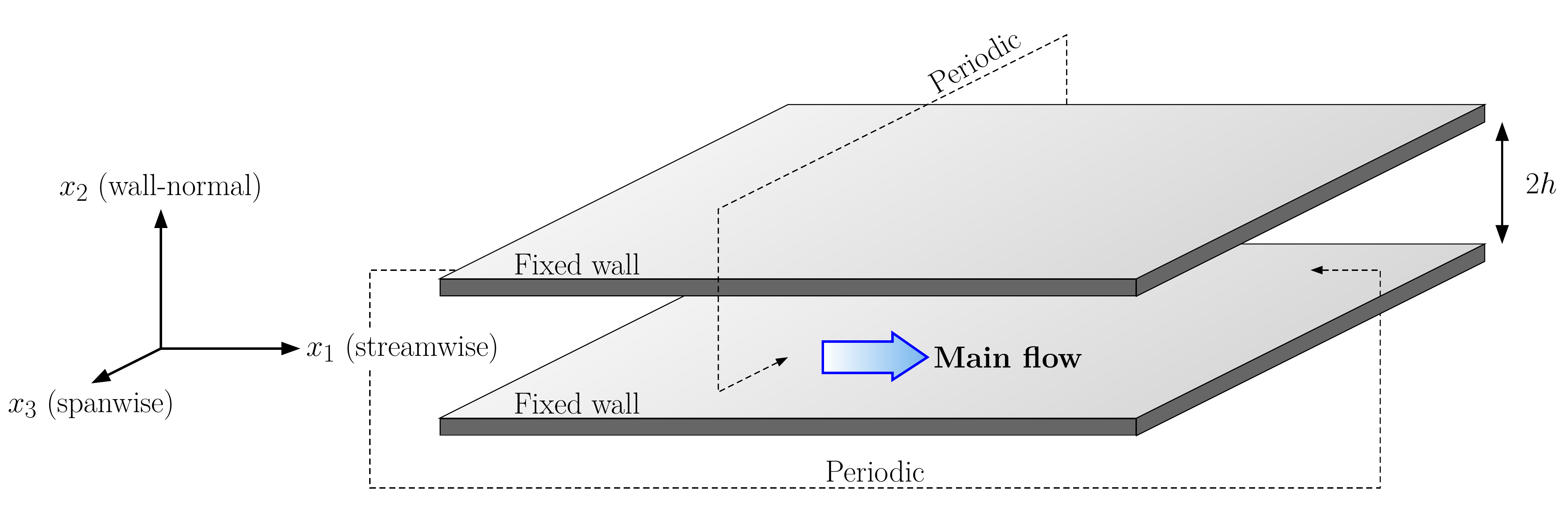}
\caption{Sketch of a turbulent channel flow configuration.}
\label{fig:sketch_channel_flow}
\end{figure}

The turbulent channel flow considered in our study refers to the flow between two parallel plates separated at a distance $2h$. The streamwise direction is assumed to be the direction $x_1\;(x)$, while the wall-normal and spanwise directions are $x_2\;(y)$ and $x_3\;(z)$, respectively. A sketch of the flow can be seen in Fig.~\ref{fig:sketch_channel_flow}. This configuration has been largely investigated and high fidelity simulations data can be found in the Refs.~\cite{moser1999, kaneda2021, hoyas2022}. Known for its simplicity, this flow is usually the academic case that researchers begin with in order to consolidate the newly developed models, including machine learning closure models proposed in recent years, see in Refs.~ \cite{fang2020, zhang2018, saezdeocarizborde2021}. Hence, we decide to thoroughly study this flow configuration in the present work.

Owing to the geometric invariance along the streamwise and spanwise directions, the velocity statistics of this flow can be considered as independent of $x_1$ and $x_3$, and thus only depend on $x_2$. Therefore, their derivatives with respect to $x_1$ and $x_3$ are all zero. The mean continuity equation in \eqref{eq:RANS} reduces to: 
\begin{equation}
    \dfrac{\partial {\overline{u}_2}}{\partial x_2}=0\label{eq:ChannelMC}
\end{equation}

\noindent which indicates that $\overline{u}_2=0$ since we have $\overline{u}_2(x_2=0, 2h)=0$ at the walls. Having also $\overline{u}_3=0$ due to the physical invariance in the $x_3$ direction and considering that the system has evolved long enough for the flow to be statistically stationary, meaning independent of time $t$, the momentum equations in (\ref{eq:RANS}) can be simplified as:
\begin{align} 
    \begin{cases}
    \text{$x_1$-direction:}&\quad 0 = -\dfrac{1}{\rho}\dfrac{\partial \overline{p}}{\partial x_1} + \nu\dfrac{\partial ^2 \overline{u}_1}{\partial x_2^2} - \dfrac{\partial \overline {u^{\prime}_1u^{\prime}_2}}{\partial x_2}\\[10pt]
    \text{$x_2$-direction:}&\quad 0 = -\dfrac{1}{\rho}\dfrac{\partial \overline{p}}{\partial x_2} - \dfrac{\partial \overline {u_2^\prime u_2^\prime}}{\partial x_2}\\[10pt]
    \text{$x_3$-direction:}&\quad 0 =  -\dfrac{\partial \overline {u^{{\prime}}_2u^{{\prime}}_3}}{\partial x_2}\\[10pt]
    \end{cases}\label{eq:RANS_channel}
 \end{align}
\noindent Here, the mean pressure gradient $\partial p/\partial x_1$ is nonzero unlike the velocity statistics, since it is indeed the driving force of the flow. 

It can be noted from above equations that only the $\overline{u^{\prime}_1u^{\prime}_2}$ closure is needed to determine the streamwise velocity profile $\overline{u}_1(x_2)$, which explains why most researchers were only interested in the predictions of $b_{12}$ component of the Reynolds stress anisotropy tensor for this flow configuration \cite{zhang2018, fang2020, saezdeocarizborde2021}. However, unlike them, we would like to fully predict the Reynolds stress anisotropy tensor in the present study since all the nonzero statistics of Reynolds stress would be practically beneficial in a real RANS simulation, especially in the near-wall region and at the beginning of the calculation in terms of convergence. Additionally, as shown in Eq.~\eqref{eq:RANS_channel}, the $b_{22}$ component is necessary to determine the pressure profile $\overline{p}(x_1, x_2)$. 

For this purpose, we will apply the aforementioned Pope's model in the case of the  turbulent channel flow. We note that the characteristics in the $x_3$ direction of this flow: $\overline{u}_3=0$ and $\dfrac{\partial }{\partial x_3}=0$, satisfy the conditions of Pope's simplified model. To write the Reynolds stress anisotropy tensor as shown in Eq.~\eqref{eq:2D_Pope}, we first give  the expressions of the normalized mean strain-rate, rotation-rate and Reynolds stress anisotropy tensors as follows:
\begin{gather}
\textbf{S}^* = \dfrac{1}{2}
\begin{bmatrix}
0 & \alpha & 0 \\
\alpha & 0 & 0 \\
0 & 0 & 0
\end{bmatrix}
\;\text{,}\;
\textbf{R}^* = \dfrac{1}{2}
\begin{bmatrix}
0 & \alpha & 0 \\
-\alpha & 0 & 0 \\
0 & 0 & 0
\end{bmatrix}
\;\text{and}\;
\textbf{b} = 
\begin{bmatrix}
\;b_{11} & b_{12} & 0 \\
\;b_{12} & b_{22} & 0 \\
\;0 & 0 & b_{33}
\end{bmatrix}   \label{eq:SRB_channel}
\end{gather} 
\noindent where $\alpha = \dfrac{k}{\epsilon}\dfrac{d\overline{u}_1}{dx_2}$ is a normalized mean velocity gradient, also the only nonzero mean velocity statistics.

Substituting Eq.~\eqref{eq:SRB_channel} into Eqs.~\eqref{eq:PopeInvariant2D} and \eqref{eq:PopeTensor2D} leads to:
\begin{equation}
\lambda_1^*=\text{tr}(\textbf{S}^{*2})=\dfrac{\alpha^2}{2} \, ,\quad \lambda_2^*=\text{tr}(\textbf{R}^{*2})=-\dfrac{\alpha^2}{2} \\\label{eq:lambdaChannel}
\end{equation}

\begin{align} 
    \begin{cases}
     	\textbf{T}^{*(0)} =  \dfrac{1}{2}\textbf{I}_2 -  \dfrac{1}{3}\textbf{I}_3&=
    	    \begin{bmatrix}
   			1/6 & 0 & 0 \\
   			0 & 1/6 & 0 \\
   			0 & 0 & -1/3 \\
   			\end{bmatrix}
            \\[10pt]   			    			
     	\textbf{T}^{*(1)} =  \textbf{S}^{*} & = 
     	  \begin{bmatrix}
   			0 & \alpha/2 & 0 \\
   			\alpha/2 & 0 & 0 \\
   			0 & 0 & 0 \\
   			\end{bmatrix}
   			\\[10pt]
      	\textbf{T}^{*(2)} =  \textbf{S}^{*}\textbf{R}^{*}-\textbf{R}^{*}\textbf{S}^{*} & = 
      	    \begin{bmatrix}
   			\alpha^2/2 & 0 & 0 \\
   			0 & -\alpha^2/2 & 0 \\
   			0 & 0 & 0 
   			\end{bmatrix}\\[10pt]
    \end{cases}\label{eq:TChannel}
\end{align}
\noindent As it is obvious that $\lambda_1^*=-\lambda_2^*$, only one invariant is relevant in our case and we will keep $\lambda_1^*$ in the following. For reminder, $\textbf{I}_2$ is taken here as $\mathrm{diag}(1,1,0)$, as required by Pope's model. From now on, we denote by $\textbf{T}^{*(03)}$ this $\textbf{T}^{*(0)}$ for referring to the location of zero in the diagonals.  

The expression of the Reynolds stress anisotropy tensor shown in Eq.~\eqref{eq:2D_Pope} can therefore be rewritten as:
\begin{equation}
\begin{split}
\textbf{b}& =g^{(0)}(\lambda_1^*)\textbf{T}^{*(0)} + g^{(1)}(\lambda_1^*)\textbf{T}^{*(1)} + g^{(2)}(\lambda_1^*)\textbf{T}^{*(2)}\\
& =g^{(0)}(\alpha)\textbf{T}^{*(0)} + g^{(1)}(\alpha)\textbf{T}^{*(1)} + g^{(2)}(\alpha)\textbf{T}^{*(2)} \label{eq:Pope_Channel}    
\end{split}
\end{equation}

Here, we have clarified the first persistent ambiguity remaining in the literature around the application of Pope's approach to the turbulent channel flow: only one invariant and three tensors are indeed necessary, and they depend merely on $\alpha$. Particularly in our domain of interest, it was found that researchers attempted to apply the general model, shown in \eqref{generalPope}, with five invariants and ten tensors, for the Reynolds stress anisotropy tensor predictions in the case of turbulent channel flow using deep neural networks \cite{fang2020}.

Another concern related to the choice of the constant tensor $\textbf{T}^{*(0)}$ has been identified. We have found that $\textbf{T}^{*(0)}$ obtained by two other possible permutations of $\textbf{I}_2$ can also form an integrity basis with $\textbf{T}^{*(1)}$ and $\textbf{T}^{*(2)}$, since we have:
\begin{equation}
\textbf{T}^{*(01)} = \mathrm{diag}(-1/3,1/6,1/6) = -\dfrac{1}{2}\textbf{T}^{*(03)} - \dfrac{1}{4\lambda_1^*}\textbf{T}^{*(2)} 
\end{equation}
and
\begin{equation}
\textbf{T}^{*(02)} = \mathrm{diag}(1/6,-1/3,1/6) = -\dfrac{1}{2}\textbf{T}^{*(03)} + \dfrac{1}{4\lambda_1^*}\textbf{T}^{*(2)} 
\end{equation}

Substituting Eq.~\eqref{eq:TChannel} into Eq.~\eqref{eq:Pope_Channel}, we obtain the expression of the Reynolds stress anisotropy tensor components as follows, the three systems of equations using $\textbf{T}^{*(01)}$, $\textbf{T}^{*(02)}$ and $\textbf{T}^{*(03)}$, respectively:
\begin{align} 
    \begin{cases}
    b_{11} = -\dfrac{1}{3}g^{(0)} - \dfrac{\alpha^2}{2}g^{(2)} \\[10pt]
    b_{12} = \dfrac{\alpha}{2}g^{(1)} \\[10pt]
    b_{22} = \dfrac{1}{6}g^{(0)} + \dfrac{\alpha^2}{2}g^{(2)} \\[10pt]
    b_{33} = \dfrac{1}{6}g^{(0)}
    \end{cases} & \quad \text{or} 
    \begin{cases}
    b_{11} = \dfrac{1}{6}g^{(0)} - \dfrac{\alpha^2}{2}g^{(2)}\\[10pt]
    b_{12} = \dfrac{\alpha}{2}g^{(1)} \\[10pt]
    b_{22} = -\dfrac{1}{3}g^{(0)} + \dfrac{\alpha^2}{2}g^{(2)} \\[10pt]
    b_{33} = \dfrac{1}{6}g^{(0)}
    \end{cases} & \text{or} 
    \begin{cases}
    b_{11} = \dfrac{1}{6}g^{(0)} - \dfrac{\alpha^2}{2}g^{(2)} \\[10pt]
    b_{12} = \dfrac{\alpha}{2}g^{(1)} \\[10pt]
    b_{22} = \dfrac{1}{6}g^{(0)} + \dfrac{\alpha^2}{2}g^{(2)} \\[10pt]
    b_{33} = -\dfrac{1}{3}g^{(0)}
    \end{cases}  \label{eq:bijPope}       
\end{align}

\begin{figure}[ht]
\centering
\includegraphics[width=0.8\textwidth]{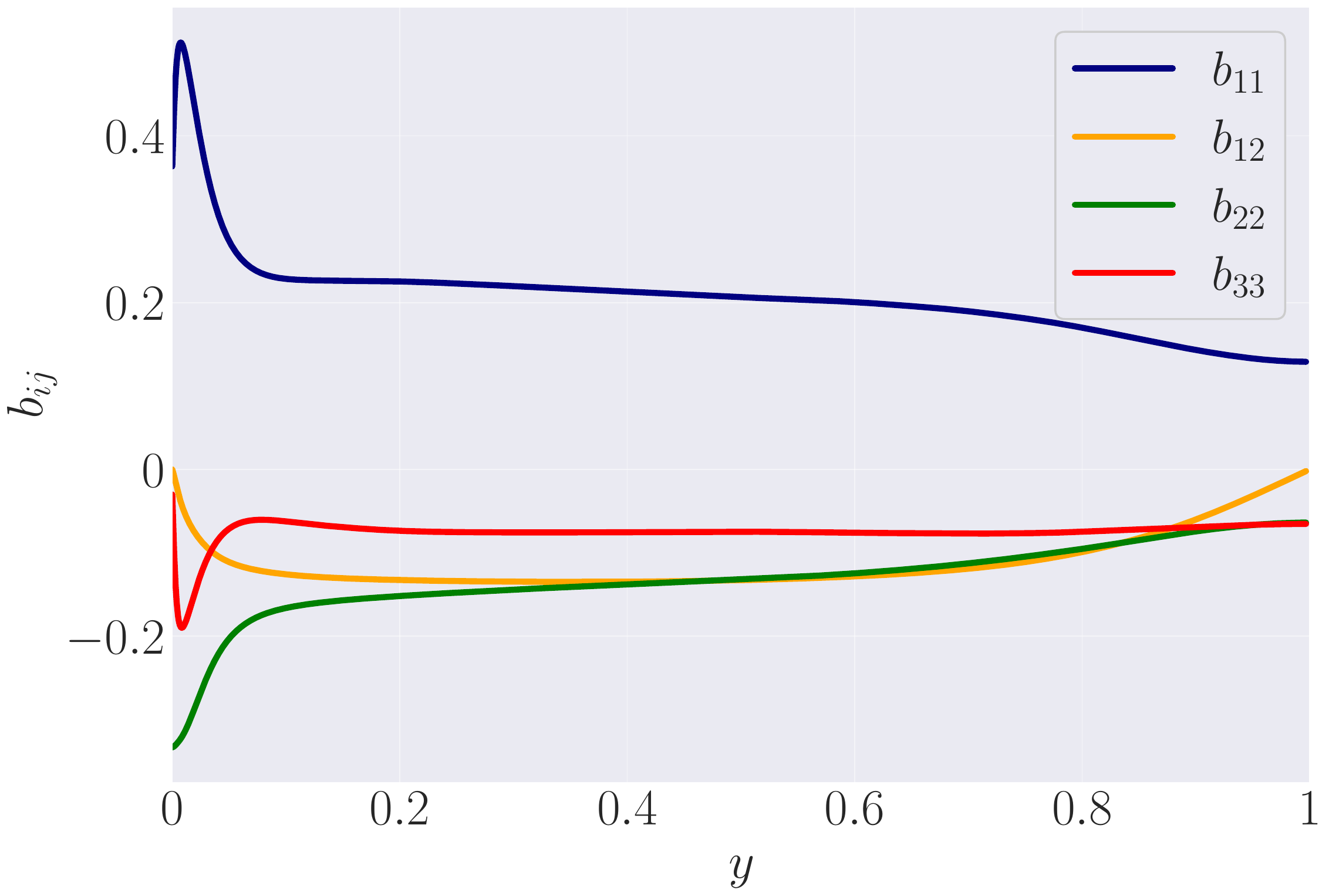}
\caption{Reynolds stress anisotropy tensor $b_{ij}$ as a function of wall distance $y$ at $\mathrm{Re}_\tau=\np{1000}$.}
\label{fig:b_diag}
\end{figure}

An arising question then concerns an optimal choice of $\textbf{T}^{*(0)}$, among these three alternatives, in the context of statistical learning. In Fig.~\ref{fig:b_diag}, we plot each nonzero component of the Reynolds stress anisotropy tensor as a function of the wall distance $y$, at the friction Reynolds number $\mathrm{Re}_\tau=\dfrac{u_{\tau}h}{\nu}=\np{1000}$, using the DNS data of \cite{moser1999}. Here, $u_{\tau}=\sqrt{\nu \dfrac{d\overline{u}_1}{dx_2}}$ is the friction velocity. Very interestingly, we may observe that $b_{12}\approx0$ and $b_{22}\approx b_{33}\approx -b_{11}/2$ at the channel center, which also holds for other DNS data at different $\mathrm{Re}_\tau$. It turns out that only $\textbf{T}^{*(01)}$ is proportional to the Reynolds stress anisotropy tensor at the channel center, and physically makes sense by including it in the basis tensors. This finding is clearly contradictory with the choice of $\textbf{T}^{*(0)}=\textbf{T}^{*(03)}$ in Pope's statement. Despite the number of literature studies applying Pope's model to the turbulent channel flow, this issue has never been discussed to our knowledge. To this end, we would like to examine this problem and propose a new generalized $\textbf{T}^{*(0)}$ hereinafter.

\subsubsection{Generalized $\textbf{T}^{*(0)}$}\label{subsec:generalizedT0}
In order not to choose arbitrarily one of the three possible expressions for $\textbf{T}^{*(0)}$, we propose here a new form of $\textbf{T}^{*(0)}$, which is aptly generalized as a linear combination of each alternative of $\textbf{T}^{*(0)}$: 
\begin{equation}
\textbf{T}^{*(0)}_{\text{gen}}=g_{01}\textbf{T}^{*(01)}+g_{02}\textbf{T}^{*(02)}+g_{03}\textbf{T}^{*(03)} \label{eq:newT0_g0}
\end{equation}
\noindent where $g_{01}$, $g_{02}$ and $g_{03}$ are coefficient functions depending on $\alpha$, instead of some fixed constants, in order to make the generalization as broad as possible under Pope's framework.

A more compact formulation of Eq.~\eqref{eq:newT0_g0} can be obtained as:
\begin{equation} 
    \begin{split}
\textbf{T}^{*(0)}_{\text{gen}}& =
     	    \begin{bmatrix}
   			-\dfrac{1}{3} g_{01} + \dfrac{1}{6} g_{02} + \dfrac{1}{6} g_{03} & 0 & 0 \\
   			0 & \dfrac{1}{6} g_{01} - \dfrac{1}{3} g_{02} + \dfrac{1}{6} g_{03} & 0 \\
   			0 & 0 & \dfrac{1}{6} g_{01} + \dfrac{1}{6} g_{02} - \dfrac{1}{3} g_{03} \\
   			\end{bmatrix}\\
   			 & = 
     	    \begin{bmatrix}
   			f_{01} & 0 & 0 \\
   			0 & f_{02} & 0 \\
   			0 & 0 & f_{03} \\
   			\end{bmatrix}\label{eq:newT0}
   	\end{split}
\end{equation}   	
where $f_{01}$, $f_{02}$ and $f_{03}$ are functions of $\alpha$ as $g_{01}$, $g_{02}$ and $g_{03}$ are, with $f_{01}+f_{02}+f_{03}=0$ to preserve the zero-trace of the Reynolds stress anisotropy tensor.

Meanwhile, we notice that the information of $\alpha$ in $\textbf{T}^{*(2)}$ is also included in $\textbf{T}^{*(0)}_{\text{gen}}$. We can therefore get rid of $\textbf{T}^{*(2)}$ and reduce the basis tensors to only $\textbf{T}^{*(0)}_{\text{gen}}$ and $\textbf{T}^{*(1)}$. Using the tensor basis composed of $\textbf{T}^{*(0)}_{\text{gen}}$ and $\textbf{T}^{*(1)}$ in the case of turbulent channel flow, we re-form Eq.~\eqref{eq:Pope_Channel} as:
\begin{equation}
\textbf{b}=\textbf{T}^{*(0)}_{\text{gen}}(\alpha) + g^{(1)}(\alpha)\textbf{T}^{*(1)} \label{eq:Pope_Channel_gen}    
\end{equation}
which can be developed into the following system of equations, giving the expression of $\textbf{T}^{*(0)}_{\text{gen}}$ shown in Eq.~\eqref{eq:newT0}:
\begin{align} 
    \begin{cases}
    b_{11} = f_{01} \\[10pt]
    b_{12} = \dfrac{\alpha}{2}g^{(1)} \\[10pt]
    b_{22} = f_{02}\\[10pt]
    b_{33} = - (f_{01}+f_{02})
    \end{cases} \label{eq:bijF}  
\end{align}

Hence, we have got in total four representations of the Reynolds stress anisotropy tensor, shown in Eqs.~\eqref{eq:bijPope} and \eqref{eq:bijF}, respectively, using either one of the three constant $\textbf{T}^{*(0)}$, or the newly proposed $\textbf{T}^{*(0)}_{\text{gen}}$. In the following, we are going to learn each of these representations of Pope's model and check the possible superiority of the $\textbf{T}^{*(0)}_{\text{gen}}$.

\subsection{Review of neural networks for turbulent channel flow}
\begin{figure}[h]
\centering
\sbox{\measurebox}{%
\begin{minipage}[b]{.5\textwidth}
\centering
\subfloat[TBNN]
  {\label{fig:TBNN_Fang}\includegraphics[height=3cm]{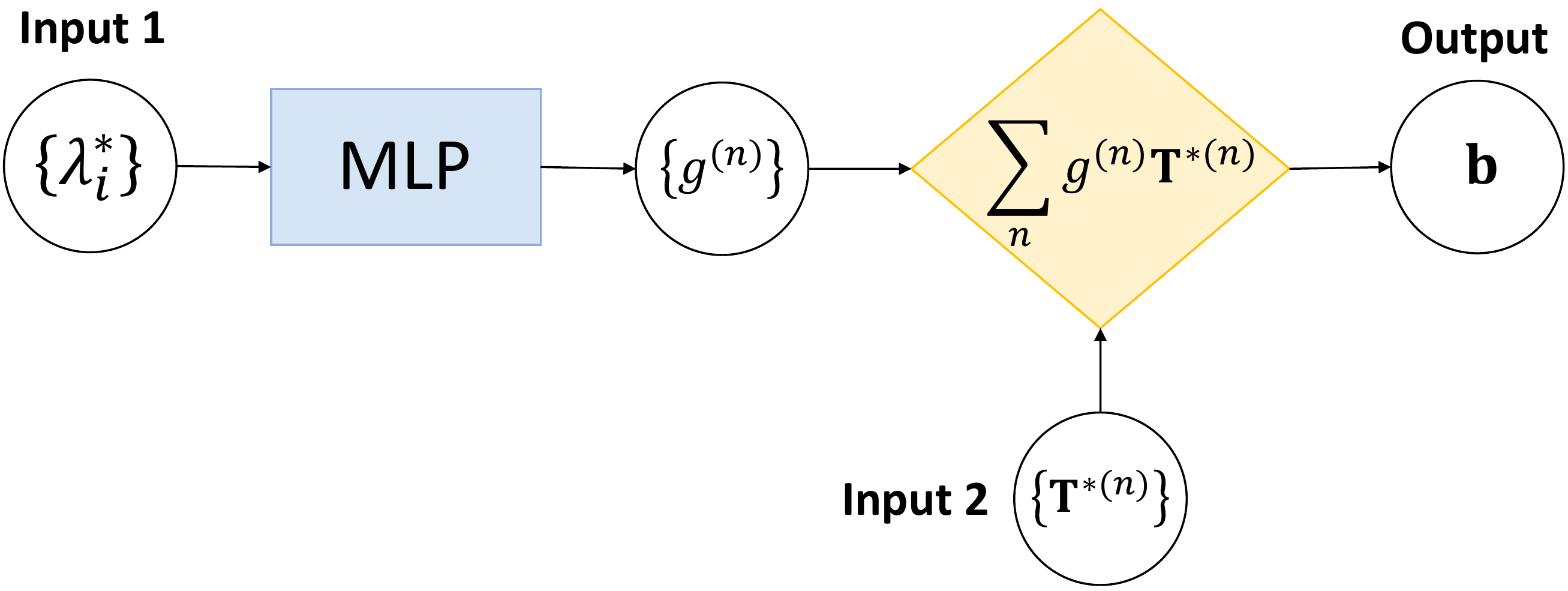}}
  
\subfloat[MLP-BC-$\mathrm{Re}_\tau$]
  {\label{fig:FCFF_Fang}\includegraphics[height=3cm]{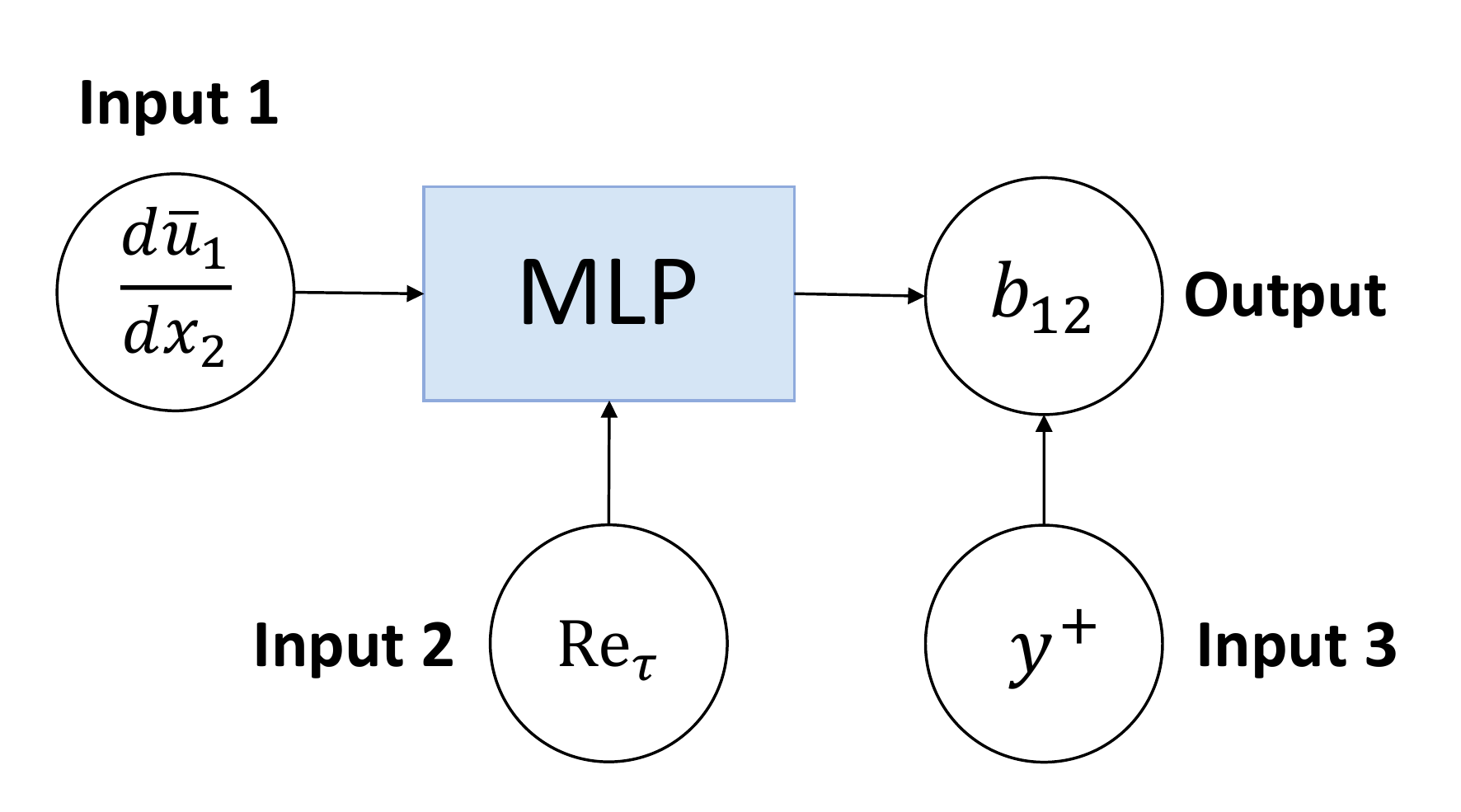}}

\subfloat[CNN-BC-$\mathrm{Re}_\tau$]
  {\label{fig:CNN_Borde}\includegraphics[height=3cm]{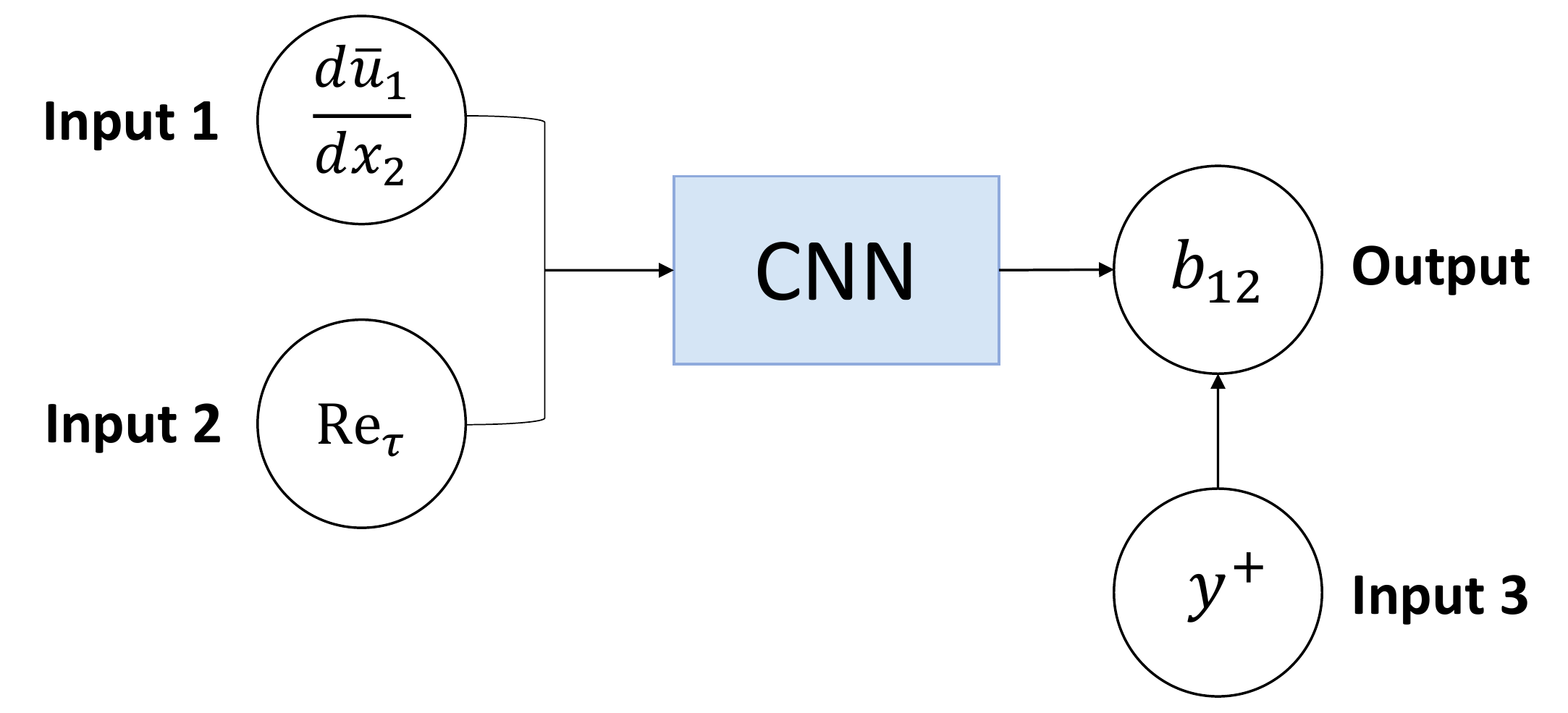}}
\end{minipage}}

\usebox{\measurebox}\qquad
  \begin{minipage}[b][\ht\measurebox][s]{.12\textwidth}
  \subfloat[CNN]
    {\label{fig:CNN}\includegraphics[height=10.28cm]{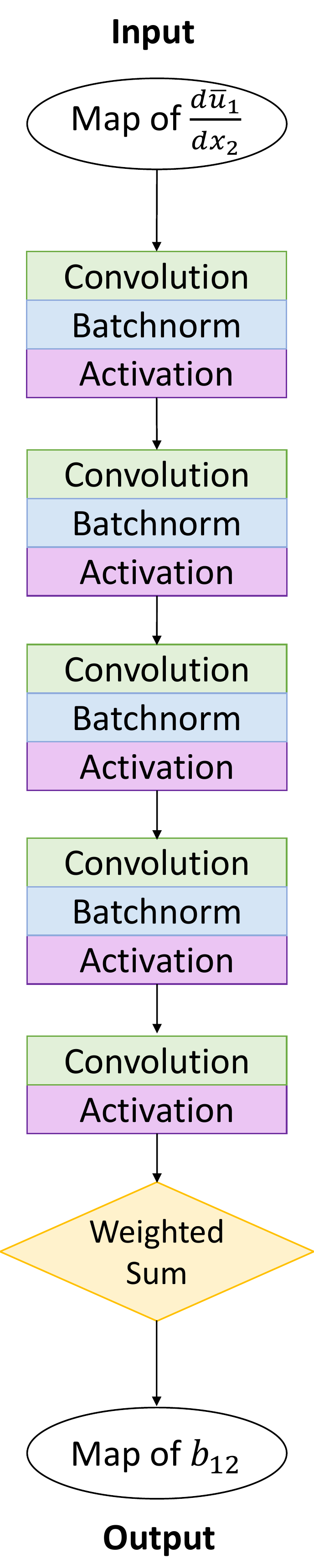}}
\end{minipage}
\caption{Diagrams of different deep learning architectures for Reynolds stress tensor predictions in turbulent channel flow, adapted from Refs.~\cite{ling2016, fang2020,saezdeocarizborde2021}.}
\label{fig:models}
\end{figure}

Based on Pope's model, Ling \textit{et al.} \cite{ling2016} designed the Tensor Basis Neural Network (TBNN), whose architecture is illustrated in Fig.~\ref{fig:TBNN_Fang}, showing a deep learning view of the Pope's general model with 5 invariants and 10 tensors. Two input layers are provided in the TBNN, one containing the invariants $\lambda_1, ..., \lambda_5$ and the other composed of the tensors $T^{(n)}$ for $n=1, ..., 10$. The former one is followed by 8 hidden layers, with 30 nodes per layer, in order to learn the ten coefficients functions $g^{(n)}$ for $n=1, ..., 10$ in the final hidden layer. This layer is then merged with the basis tensors input layer by element-wise multiplications so as to give final predictions on Reynolds stress anisotropy tensor. This innovative architecture guarantees Galilean invariance and rotational invariance as Pope's model does. 

Ling \textit{et al.} \cite{ling2016} trained, validated and tested the neural network on a high-fidelity database of nine flow cases. These flow cases represent a wide variety of canonical flows, varying from duct flows, channel flows, a jet in a cross-flow for training, to a wall-mounted cube in a cross-flow for validation and flow over a wavy wall for test. Despite the flow diversity under consideration, both a priori predictions on Reynolds stress anisotropy tensor and a posteriori results on mean velocity yielded by TBNN were more accurate than traditional RANS models and a generic neural network that did not embed invariance properties.

A series of studies has been conducted with the inspiration of above work. Zhang \textit{et al.} \cite{zhang2018} simplified the TBNN for the case of turbulent channel flow to predict only one component in the Reynolds stress anisotropy tensor, the $b_{12}$. A smaller baseline network structure with 4 hidden layers and 20 nodes per layer was used in order to avoid over-fitting. Three methods were applied upon this baseline model. Regularization was introduced as the first method, while the other methods focused on variable selection: one embedded the dimensionless wall distance $y^+=\dfrac{yu_\tau}{\nu}$ in the input for better predictions in the near-wall region, and the other used only 2 invariants and 3 tensors by considering the 2D nature of turbulent channel flow in a RANS modeling framework. All these models were trained with Moser \textit{et al.}'s DNS database \cite{moser1999} for channel flows at four turbulent Reynolds numbers, among which the data set at $\mathrm{Re}_\tau=\np{2000}$ was only used for testing. These simplified and adapted TBNN model outperformed the original model by comparing the $b_{12}$ predictions.

Recently, other deep learning architectures than TBNN were proposed for turbulent channel flow configuration. Fang \textit{et al.} \cite{fang2020} directly employed the Fully-Connected Feed-Forward (FCFF) neural network, also known as Multi-Layer Perceptron (MLP), with 5 hidden layers and 50 nodes per layer. The normalized velocity gradient was used as a basis input to predict $b_{12}$. They attempted to incorporate more physical information in the model. They obtained the best performance by embedding in the model a constraint function depending on $y^+$, in order to enforce the no-slip boundary condition at the channel wall, and the $\mathrm{Re}_\tau$ in the input feature, which also outperformed the TBNN. A diagram of their model can be seen in Fig.~\ref{fig:FCFF_Fang}. The significance of non-local effects was also emphasized in their work, but the results were not as expected while considering these effects in their MLP framework. Sáez de Ocáriz Borde \textit{et al.} \cite{saezdeocarizborde2021} expanded this work and designed a Convolutional Neural Network (CNN) architecture, with 5 one dimensional convolution layers as illustrated in Fig.~\ref{fig:CNN}, in order to better capture non-local effects for the prediction of $b_{12}$. Unlike earlier models which gave one-point predictions, the newly proposed CNN takes an array containing all the normalized velocity gradients into inputs and predicts an array of $b_{12}$ at all vertical locations. The same boundary condition enforcement and friction Reynolds number incorporation techniques proposed by Fang \textit{et al.} \cite{fang2020} were used, as shown in Fig.~\ref{fig:CNN_Borde}. This model yielded even better results by evaluating the $R^2$ score. 

Despite aforementioned successes in predicting the Reynolds stress anisotropy tensor in the case of turbulent channel flow, there is still a lack of physical explanation on the input feature selection. We would like to underline that in today's physics-based machine learning, a deeper scientific understanding is required not just to understand why our models fail, but also to understand why they succeed, as a future reference. For this sake, we are going to extent former study upon simpler MLP architectures with different combinations of input features formerly used, in order to understand the role of each feature in the neural networks. Meanwhile, having previously clarified some persisting ambiguities concerning the application of Pope's model for turbulent channel flow in Section \ref{subsec:TCF}, we will re-examine the performance of the TBNN model for this specific case and propose the augmented TBNN architectures with additional input features other than Pope's representation. We will also investigate the sensitivity of the performance of augmented TBNN on the choice of $\textbf{T}^{*(0)}$ and compare their predictive capacity with generic MLP architectures. Beyond these extensions, unlike previous study on turbulent channel flow focusing only on $b_{12}$ component, we aim at providing predictions for all terms with nonzero statistics, that is $b_{11}, b_{12}, b_{22}$ and $b_{33}$. Last but not least, models in previous studies were only evaluated at one turbulent Reynolds number $\mathrm{Re}_\tau$ at a time, either in an interpolation case, which means that the tested $\mathrm{Re}_\tau$ is within the range of the learning database, or on the contrary in an extrapolation case. In the present study, thanks to newly available DNS databases \cite{kaneda2021, hoyas2022}, we will for the first time be able to assess both interpolating and extrapolating predictability of our neural networks simultaneously. This is challenging but critical since the prediction model should be accurate in both scenarios for practical use. 

\section{Methodology}\label{sec:method}
\subsection{Data set}
\begin{figure}[ht]
\centering
\includegraphics[scale=0.3]{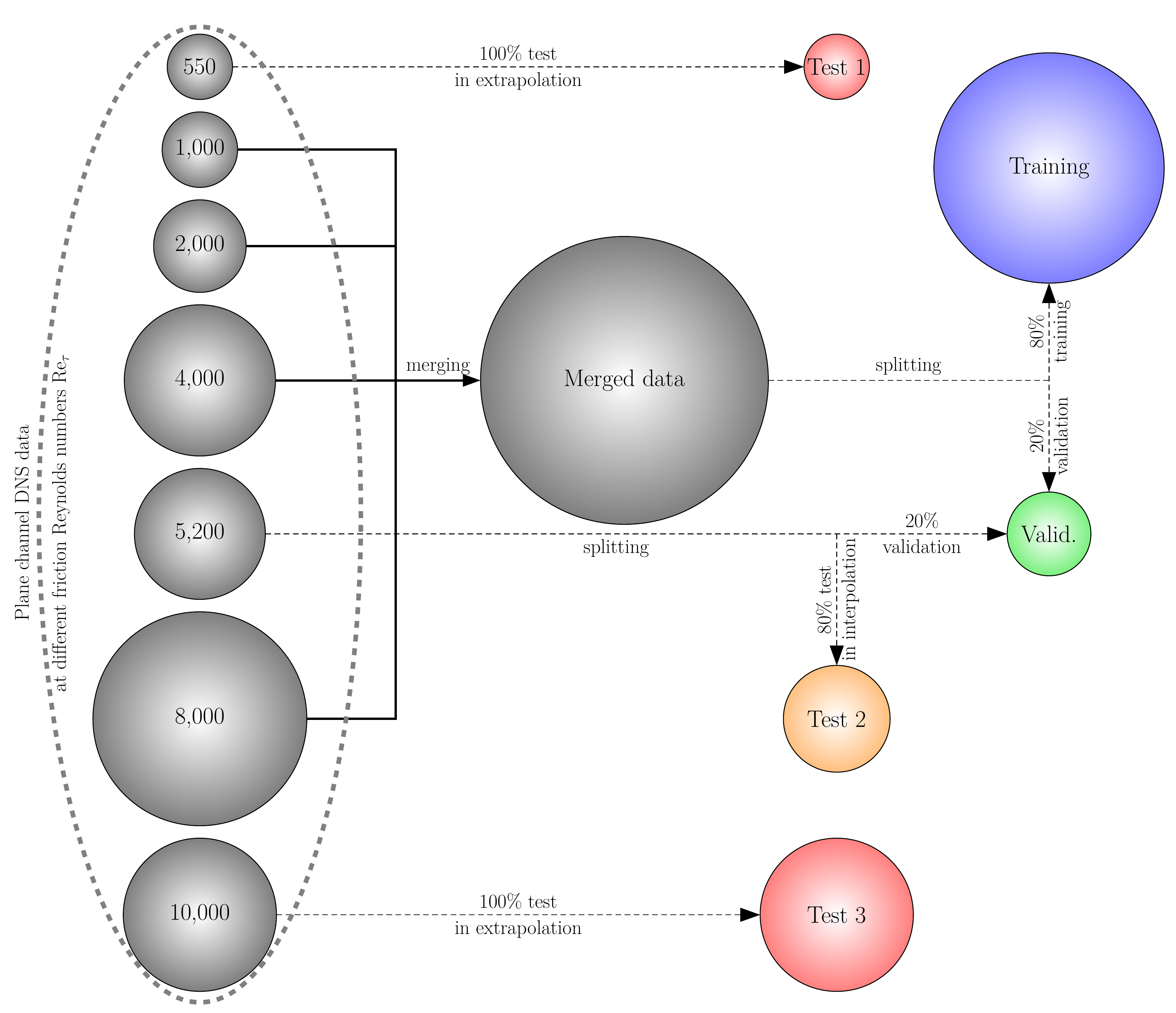}
\caption{Diagram of data split process. The size of each bubble is proportional to the data size.}
\label{fig:data_split}
\end{figure}

The data set used in the present work consists of DNS data at seven different friction Reynolds numbers $\mathrm{Re}_\tau=[550; \np{1000}; \np{2000}; \np{4000}; \np{5200}; \np{8000}; \np{10000}]$ for turbulent channel flow \cite{moser1999, kaneda2021, hoyas2022}. Among which, data at $\mathrm{Re}_\tau=[550; \np{10000}]$ were only used in test set, data at $\mathrm{Re}_\tau=\np{5200}$ were split randomly into 80\% test data and 20\% validation data. The remaining data were then divided randomly into 80\% training data and 20\% validation data. The test set here was used to evaluate the predictive capacity of our neural network models, as said before, both for interpolation and for extrapolation. A summary of the size for each data set can be seen in Table~\ref{tab:dataset}. An illustration of the data split process is shown in Fig.~\ref{fig:data_split}.

We identify here three input features, the normalized parameter $\alpha$, which is the single parameter that $b_{ij}$ depends on according to Pope's model applied to turbulent channel flow, as detailed previously in Section \ref{subsec:TCF}; the dimensionless wall distance $y^+$ and turbulent Reynolds number $\mathrm{Re}_\tau$. The reasons for choosing these features will be explained in Section \ref{sec:feature_visu}. Our learning targets are all the nonzero components of the Reynolds stress anisotropy tensor $\textbf{b}$, namely $b_{11}$, $b_{12}$, $b_{22}$, and $b_{33}$.

\begin{table*}
\begin{ruledtabular}
\centering
\caption{\label{tab:dataset}Data size at each friction Reynolds number.}
\begin{tabular}{lccccccc}
$\mathrm{Re}_\tau$ & 550 & \np{1000} & \np{2000} & \np{4000} & \np{5200} & \np{8000} & \np{10000} \\\hline
Data size & 191 & 255 & 383 & \np{1023}  & 767 & \np{2047} & \np{1050} \\
Reference & \cite{moser1999}  & \cite{moser1999} & \cite{moser1999} & \cite{kaneda2021} & \cite{moser1999} & \cite{kaneda2021} & \cite{hoyas2022} \\
\end{tabular}
\end{ruledtabular}
\end{table*}

\subsubsection{Pre-processing}
\begin{figure}[ht]
  \centering
  \begin{subfigure}{\linewidth}
    \centering
    \includegraphics[width=\textwidth]{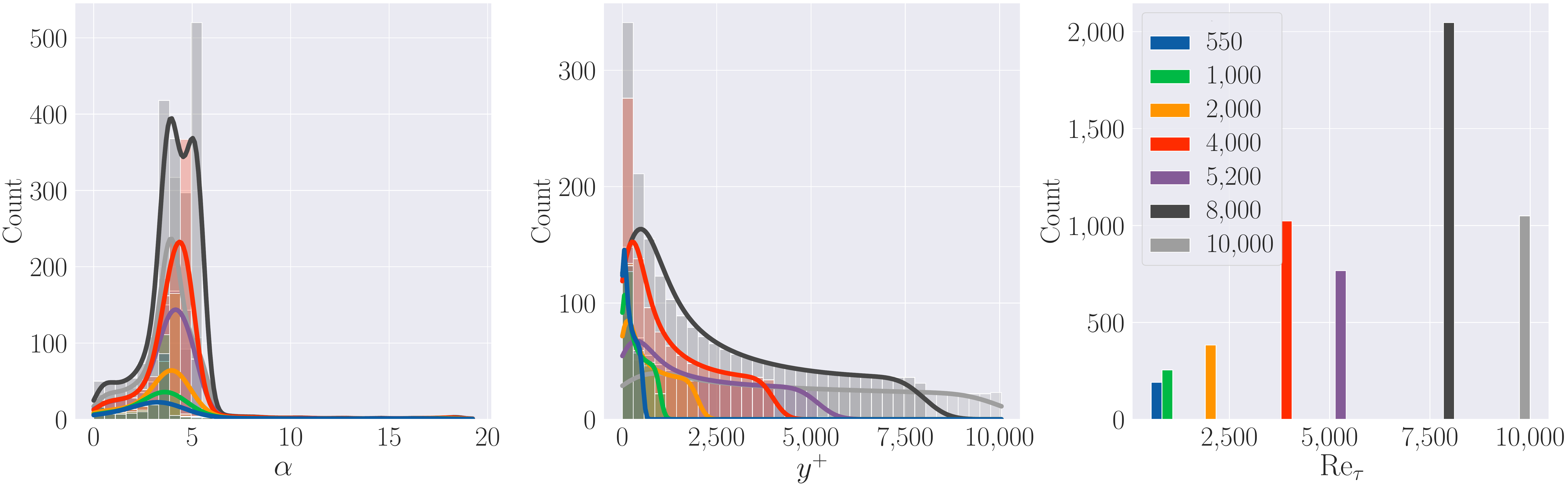}
    \caption{Before pre-processing.}
    \label{fig:inputs_before}
  \end{subfigure}

  \begin{subfigure}{\linewidth}
    \centering
    \includegraphics[width=\textwidth]{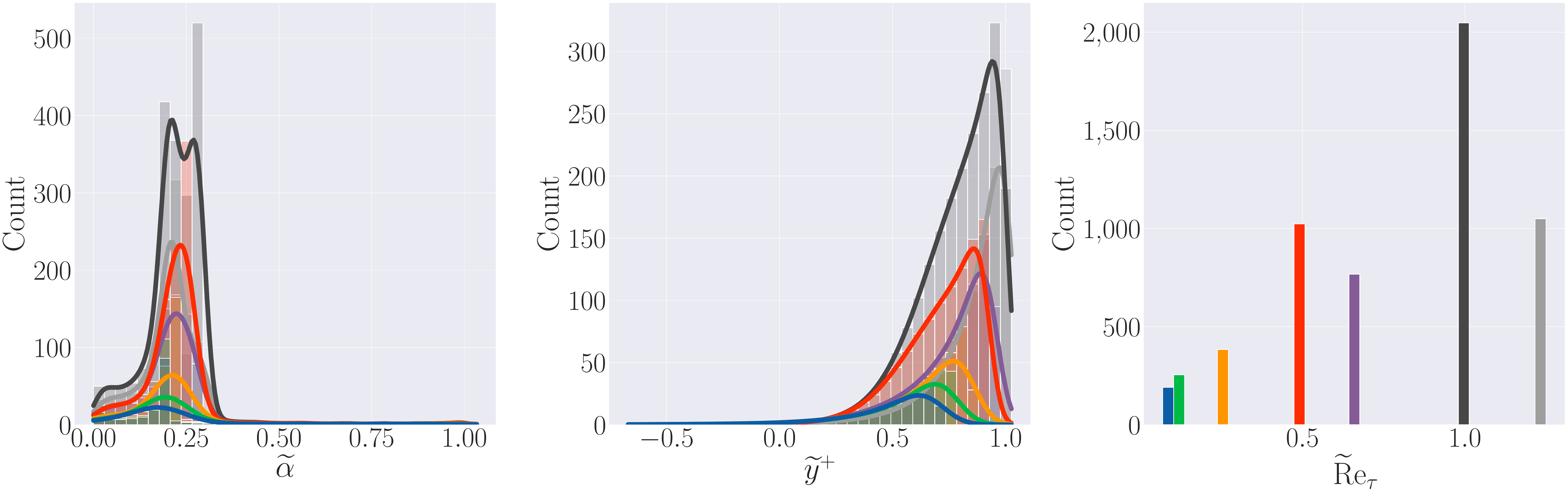}
    \caption{After pre-processing.}
    \label{fig:inputs_after}
  \end{subfigure}  
  \caption{Distribution of the input features before and after pre-processing, for various DNS experiments with different $\mathrm{Re}_\tau$.}  
\end{figure}

The data quality has a significant impact on the performance of deep learning framework. During training, the neural network tends to assign more weight to the input with a larger range of values, especially if there is noticeable difference among their scales. This is not an ideal scenario since other smaller inputs may be also important for the predictions. To avoid this scenario, we firstly visualized the distribution of our original training data, shown in Fig.~\ref{fig:inputs_before}. We remark obviously two problems from this visualization: (1) the range of $\alpha$ is much smaller than that of $y^+$ and $\mathrm{Re}_\tau$; (2) the distribution and the range of $y^+$ differ from one $\mathrm{Re}_\tau$ to another. In fact, the latter is caused by the definition of $y^+$, from which we have $\mathrm{Re}_\tau = \mathrm{max} (y^+)$.

Hence, it is necessary to pre-process our input data before feeding them into the model, in order to increase the robustness of our neural network training. A common normalization technique named max normalization is used, which aims to divide a feature by its maximum. This normalization is directly applied on $\alpha$ and $\mathrm{Re}_\tau$, while for $y^+$, a log-transformation is preliminarily performed to address the long tail issue. Accordingly, the pre-processed input features, $\widetilde{\alpha}$, $\widetilde{y}^+$ and $\widetilde{\mathrm{R}}\mathrm{e}_{\tau}$, are as follows:
\begin{equation}
\widetilde{\alpha}=\dfrac{\alpha}{\mathrm{max} (\alpha)}
\end{equation}
\begin{equation}
\widetilde{y}^+=\dfrac{\mathrm{log}(y^+)}{\mathrm{max} (\mathrm{log}(y^+)}
\end{equation} 
\begin{equation}
\widetilde{\mathrm{R}}\mathrm{e}_{\tau}=\dfrac{\mathrm{Re}_\tau}{\mathrm{max} (\mathrm{Re}_\tau)}
\end{equation}
where $\mathrm{max}(x)$ refers to the maximum value of $x$ in the training set.

The distributions of the input features after pre-processing are shown in Fig.~\ref{fig:inputs_after}. Above problems have been successfully dealt with, especially when we see that the three input features are rescaled into comparable ranges. 

On the other side, it would be generally preferable to normalize separately each regression target of a neural network in order to avoid one of them dominating the loss function. However, we prefer not to do so for the purpose of preserving the zero trace of Reynolds stress tensor. A global reduction based on the Frobenius norm of $b_{ij}$ is performed instead:
\begin{equation}
\sigma_{\textbf{b}}=\sqrt{\dfrac{1}{m} \left[ \sum_{k=1}^{3} b_{kk}^2 + b_{12}^2 \right]} 
\end{equation}
\begin{equation}
\widetilde{\textbf{b}}=\dfrac{\textbf{b}}{\sigma_{\textbf{b}}}
\end{equation}
where $m$ is the number of training samples.

\subsubsection{Feature visualization}\label{sec:feature_visu}
\begin{figure}[ht]
\centering
\includegraphics[width=\textwidth]{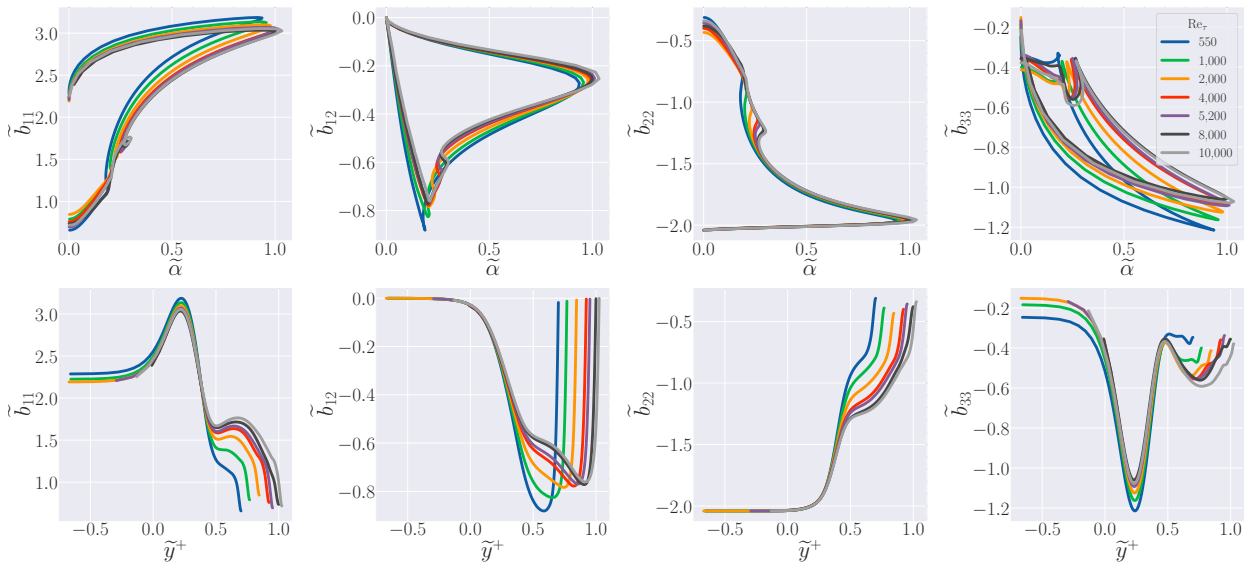}
\caption{Visualization of $\widetilde{b}_{ij}$ as a function of $\widetilde{\alpha}$ and $\widetilde{y}^+$, for various DNS experiments with different $\mathrm{Re}_\tau$.}
\label{fig:b_visu_trans}
\end{figure}

The various anisotropy components $b_{ij}$ calculated from the DNS data after global reduction, are plotted as a function of each feature after pre-processing in Fig.~\ref{fig:b_visu_trans}. This is an interesting figure, from which we can clearly observe the limitations of Pope's model revealed by the DNS data. We remind here that according to Pope's assertion applied to the turbulent channel flow, $b_{ij}$ depend merely on $\alpha$. However, we can see from the four sub-figures in the first line of Fig.~\ref{fig:b_visu_trans} the existence of the multi-value issue: $b_{ij}$ is not a function of $\alpha$, as for a given $\alpha$ value, multiple values of $b_{ij}$ are possible. Besides, we remarked from the other four sub-figures in the second line that $b_{ij}$ somewhat depends on the turbulent Reynolds number especially when $y^+$ becomes closer to its upper limit, which Pope's model does not reflect, either. 

As a result, to overcome these difficulties, it is necessary to include other representative input features into the deep neural networks in order to forecast $b_{ij}$ accurately. We propose to rely on $y^+$ and $\mathrm{Re}_\tau$. Firstly, we can see from Fig.~\ref{fig:b_visu_trans} that $b_{ij}$ are functions of $y^+$ at one given $\mathrm{Re}_\tau$; Secondly, we expect to include the $\mathrm{Re}_\tau$ as a classifier of data originating from flows with different turbulent levels. In the following study, case studies with different feature combinations will be performed for a better understanding of the optimal feature set.

\subsection{Neural networks}
We introduce in the present work two different architectures of neural networks (see Ref.~\cite{Goodfellow-et-al-2016} for details) to fully predict the Reynolds stress anisotropy tensor for turbulent channel flow. 

The first one is an augmented TBNN model specially designed for this flow configuration with only one invariant and three tensors, as previously clarified in Section \ref{subsec:TCF}. We distinguish two slightly different models: one using the three alternative constant $\textbf{T}^{*(0)}$ and the other using the newly proposed generalized $\textbf{T}^{*(0)}_{\text{gen}}$. 
As two new features are included into the augmented TBNN model apart from $\alpha$, the expression of the Reynolds stress anisotropy tensor shown in Eqs.~\eqref{eq:Pope_Channel} and \eqref{eq:Pope_Channel_gen} can be rewritten as:
\begin{equation}
\textbf{b} =g^{(0)}(\alpha, y^+, \mathrm{Re}_\tau)\textbf{T}^{*(0)} + g^{(1)}(\alpha, y^+, \mathrm{Re}_\tau)\textbf{T}^{*(1)} + g^{(2)}(\alpha, y^+, \mathrm{Re}_\tau)\textbf{T}^{*(2)}  
\end{equation}
for the former model, and
\begin{equation}
\textbf{b}=\textbf{T}^{*(0)}_{\text{gen}}(\alpha, y^+, \mathrm{Re}_\tau) + g^{(1)}(\alpha, y^+, \mathrm{Re}_\tau)\textbf{T}^{*(1)}   
\end{equation}
for the latter one.

\begin{figure}[ht]
\begin{subfigure}{.48\linewidth}
\centering
\includegraphics[width=\linewidth]{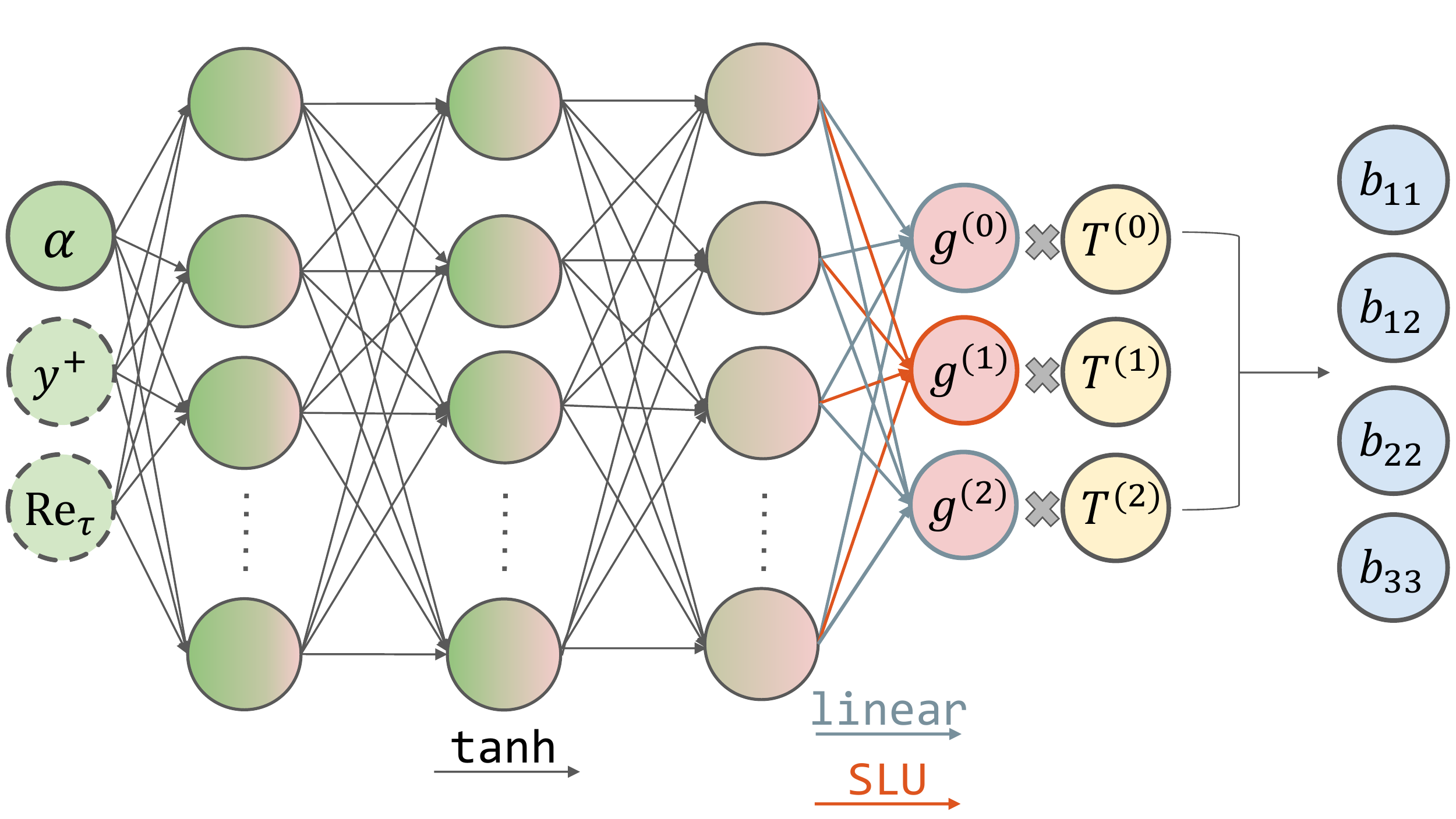}
\caption{}
\label{fig:sketch_TBNN1}
\end{subfigure}%
\begin{subfigure}{.48\linewidth}
\centering
\includegraphics[width=\linewidth]{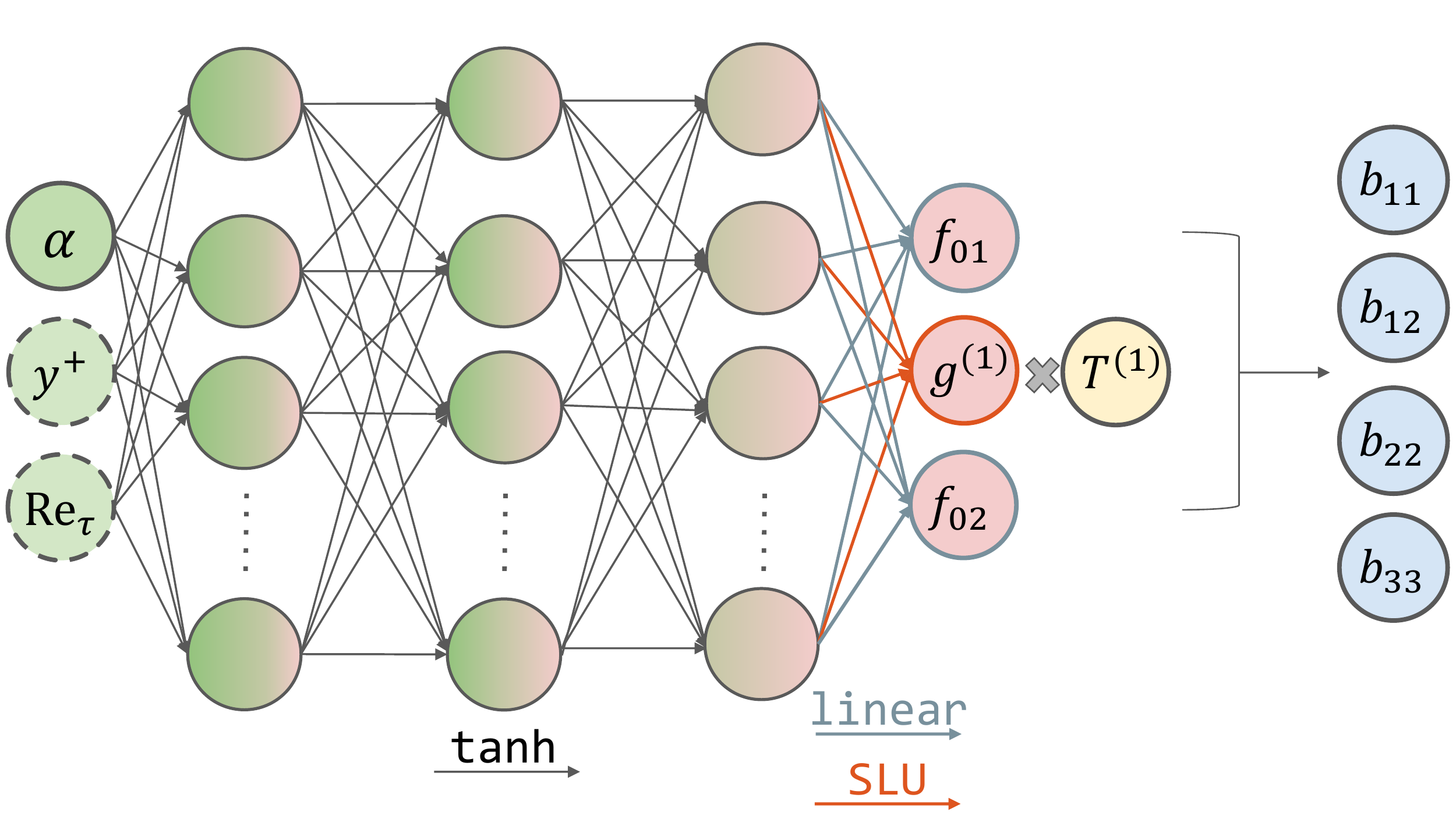}
\caption{}
\label{fig:sketch_TBNN2}
\end{subfigure}\\[1ex]
\begin{subfigure}{\linewidth}
\centering
\includegraphics[width=0.55\linewidth]{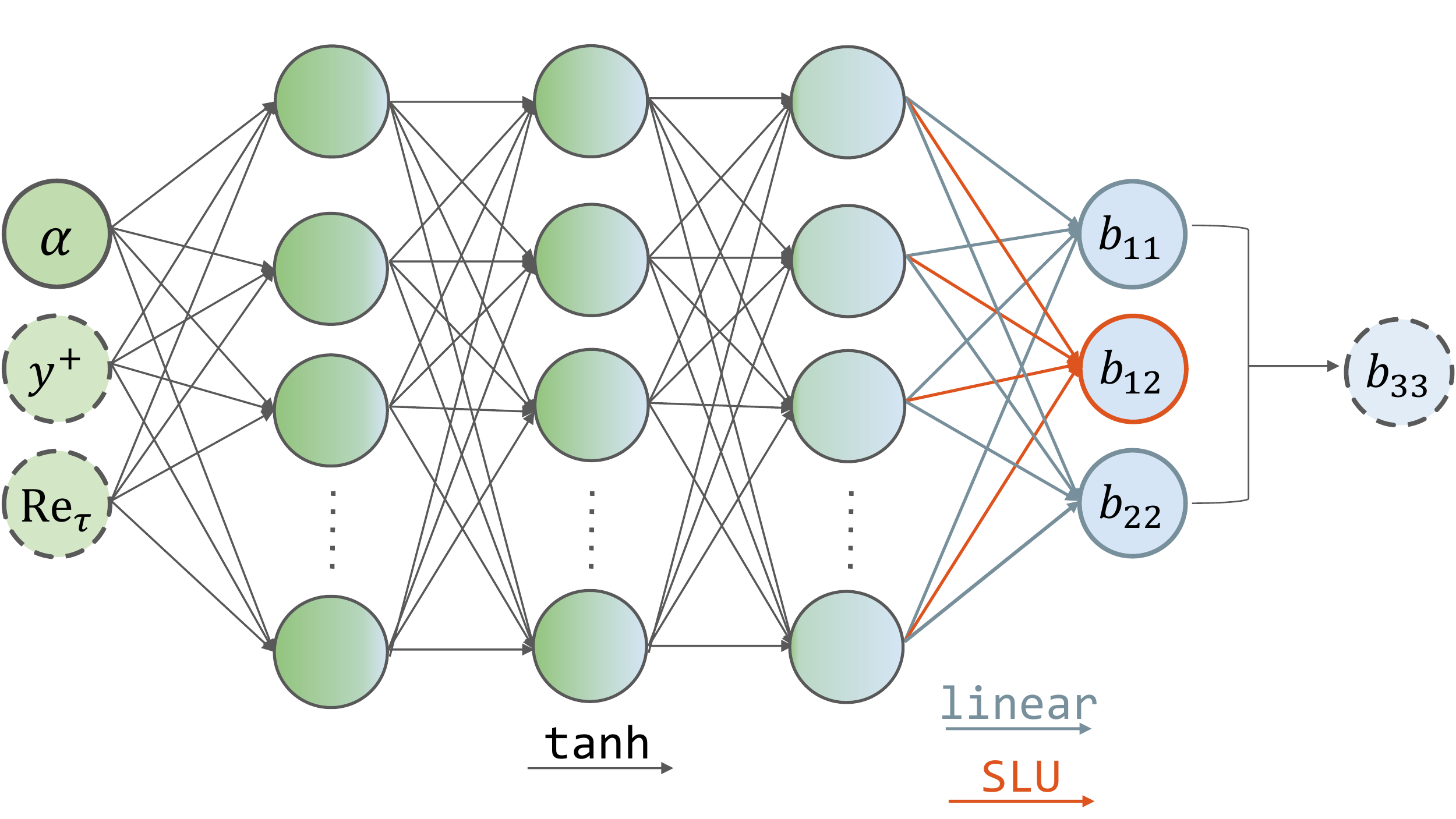}
\caption{}
\label{fig:sketch_MLP}
\end{subfigure}
\caption{Diagrams of neural network architectures used in the present work.}
\label{fig:NN}
\end{figure}

We use three hidden layers with 10 nodes per layer, which are activated by hyperbolic tangent function (tanh). The output layer contains three nodes for the three corresponding coefficient functions. Except for the output node of $g^{(1)}$ which is activated by the Softplus Linear Unit (SLU) to assure that the predicted $g^{(1)}$ is negative (see Section \ref{subsec:EVM}), the others are linearly activated by default. Illustrations of these two augmented TBNN models can be seen in Fig.~\ref{fig:sketch_TBNN1} and Fig.~\ref{fig:sketch_TBNN2}, respectively.

The second architecture is of MLP type, as illustrated in Fig.~\ref{fig:sketch_MLP}, with the same number of hidden layers and nodes as the augmented TBNN models. The outputs of this model are $b_{11}$, $b_{12}$ and $b_{22}$, while $b_{33}$ is evaluated as $-(b_{11}+b_{22})$ to guarantee the zero-trace. The activation function of the hidden layers is also tanh. Since $\alpha$ is found to be always positive, the output node of $b_{12}$ is activated by the SLU to guarantee its negativity, while the others are linearly activated. 

Above models are all implemented by calling an open-source library, named TensorFlow, in the language of Python. 
The loss function is defined as the Mean Squared Error (MSE) based on the Reynolds stress anisotropy tensor components:
\begin{equation}
\mathrm{MSE} = \dfrac{1}{4m} {\sum_{i=1}^{m}} \left[ {\sum_{k=1}^{3}} (b_{kk} - \widehat{b}_{kk}) ^2 + (b_{12} - \widehat{b}_{12}) ^2 \right]
\end{equation}
where the predicted outputs are denoted with a hat.

We use the coefficient of determination $R^2$ as evaluation metrics:
\begin{equation}
R^2 = 1 - \dfrac{\sum ({y}_{i} - \widehat{y}_{i})^2}{\sum (y_{i} - \overline{Y})^2}
\end{equation}
\noindent where $\widehat{y}_{i}$ is the predicted $i^{\text{th}}$ value, $\overline{y}_{i}$ is the actual $i^{\text{th}}$ value and $\overline{Y}$ is the mean of the true values.

A weighted $R^2$ error is defined to evaluate the global predictive performance:
\begin{equation}
R^2_{\text{test}} = \dfrac{\sum m_{\text{test},i} \times R^2_{\text{test},i}}{\sum m_{\text{test},i}}
\end{equation}
\noindent where $m_{\text{test},i}$ and $R^2_{\text{test},i}$ are respectively the size and the $R^2$ error of the $i^{\text{th}}$ test set. We note from now on test 1, 2 and 3 respectively for test set at $\mathrm{Re}_\tau=550; \np{5200}$ and $\np{10000}$. Hence, test 1 and 3 are extrapolation tests and test 2 is an interpolation test.

A summary of hyperparameter setting is shown in Table~\ref{tab:hyperpara}.

\begin{table}
\begin{ruledtabular}
\centering
\caption{\label{tab:hyperpara}Hyperparameter setting.}
\begin{tabular}{lc}
Name & Value \\\hline
Number of hidden layers & 3 \\
Number of nodes per hidden layer & 10 \\
Loss function & MSE \\
Optimization algorithm & Adam \\
Maximum learning rate & 0.0001 \\
Batch size & 64 \\
Weight initialization function & Truncated normal distribution \\
\end{tabular}
\end{ruledtabular}
\end{table}

\section{Results}\label{sec:results}
Eight case studies are performed in our work (summarized in Table~\ref{tab:case}), among which Case 1 - Case 4 are conducted to investigate the impact of input features using the more flexible MLP model, and Case 5 - Case 8 aim to find out the optimal choice of $\textbf{T}^{*(0)}$ for the augmented TBNN. Comparison between these two sets of case studies using respectively MLP and augmented TBNN allows us to quantify a better neural network model. Results will be shown in the following sections.

\begin{table}
\begin{ruledtabular}
\centering
\caption{\label{tab:case}Summary of case studies.}
\begin{tabular}{lccccccc}
Case & Model & Features & $\textbf{T}^{*(0)}$\\\hline
1 & MLP & $\{\alpha\}$ & /\\
2 & MLP & $\{\alpha, y^+\}$ & /\\
3 & MLP & $\{\alpha, \mathrm{Re}_\tau\}$ & /\\
4 & MLP & $\{\alpha, y^+, \mathrm{Re}_\tau\}$ & /\\
5 & Augmented TBNN & $\{\alpha, y^+, \mathrm{Re}_\tau\}$ & $\textbf{T}^{*(01)}$ \\
6 & Augmented TBNN & $\{\alpha, y^+, \mathrm{Re}_\tau\}$ & $\textbf{T}^{*(02)}$ \\
7 & Augmented TBNN & $\{\alpha, y^+, \mathrm{Re}_\tau\}$ & $\textbf{T}^{*(03)}$ \\
8 & Augmented TBNN & $\{\alpha, y^+, \mathrm{Re}_\tau\}$ & $\textbf{T}^{*(0)}_\mathrm{gen}$ \\
\end{tabular}
\end{ruledtabular}
\end{table}

\subsection{Feature selection}
We train the MLP model with different feature combinations in order to figure out the role of each entry and find the ideal set: $\{\alpha\}$ in Case 1, $\{\alpha, y^+\}$ in Case 2, $\{\alpha, \mathrm{Re}_\tau\}$ in Case 3 and  $\{\alpha, y^+, \mathrm{Re}_\tau\}$ in Case 4. The training is respectively stopped at 1,079, 19,940, 2,455 and 5,217 epochs for each case when the loss value evaluated on validation data set starts to stagnate.

\begin{figure}[ht]
\centering
\includegraphics[scale=0.4]{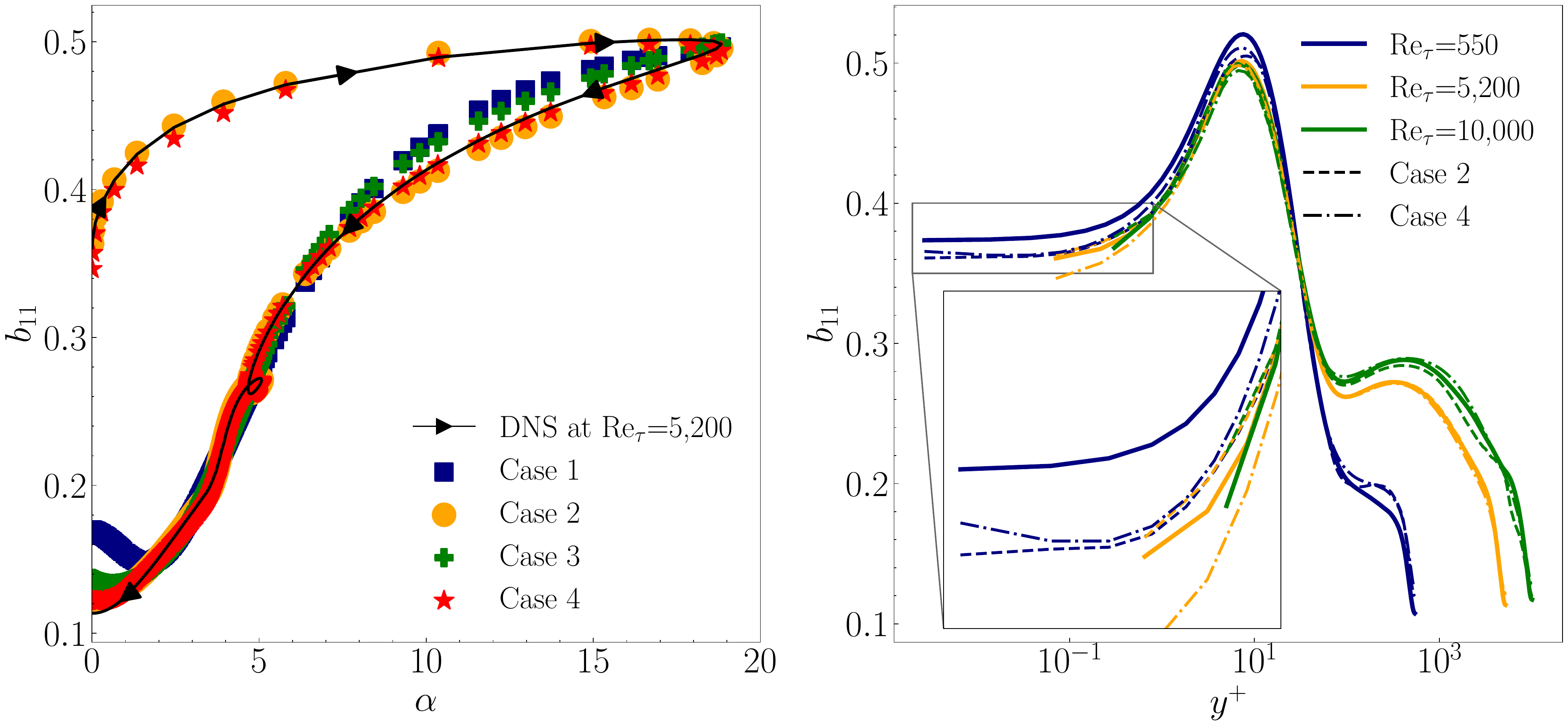}
  \caption{$b_{11}$ predictions compared with DNS data (in solid lines, pointing from the wall to the center of the channel). Left: at $\mathrm{Re}_\tau=\np{5200}$ for Case 1 - Case 4; right: at $\mathrm{Re}_\tau=550; \np{5200}; \np{10000}$ for Case 2 (in dashed lines) and Case 4 (in dotted lines).}
\label{fig:b11_Case1-4}
\end{figure}

Fig.~\ref{fig:b11_Case1-4} shows the $b_{11}$ predictions in Case 1 - Case 4 for the test 2 with $\mathrm{Re}_\tau=\np{5200}$, compared with the DNS data. Noticeably, Case 1 and Case 3 have similar behavior and both completely fail to predict the upper branch of $b_{11}$, corresponding to the near-wall region. The failure of Case 1 confirms the limitation of Pope's model, which relates each $b_{ij}$ component only to $\alpha$ for turbulent channel flow. Based on such assumption, the neural network built in Case 1 tries to construct a function between $\alpha$ and each learning target $b_{ij}$, which is indeed not realizable according to the visualization performed in Section \ref{sec:feature_visu}. 

On the other hand, by including $y^+$ into the feature set, Case 2 and Case 4 are able to confront the multi-valued issue and to capture the trends and behavior of DNS data with only minor discrepancies, as shown in Fig.~\ref{fig:b11_Case1-4} on the left side. Results given in these two cases closely overlap and are in good agreement with the DNS data. To gain a deeper insight into the difference between them, we plot $b_{11}-y^+$ curves predicted for all the test sets  on the right side of Fig.~\ref{fig:b11_Case1-4}. Despite the good performance shown in Case 2, we particularly find that it gives same predictions for all of the three turbulent Reynolds numbers in the near-wall region, which can be observed more clearly in the zoomed-out window. This makes sense because the magnitudes of $y^+$ are identically small in the near-wall region (corresponding to the region with $y^+ < 10$ in the viscous sublayer) and vary from one $\mathrm{Re}_\tau$ to another while approaching to the center of the channel. More precisely, the lower limit of the $y^+$ is always set at the order of $0.1$ for flows at all the Reynolds numbers in order to guarantee the resolution of a given DNS experiment, while by definition its upper limit is the friction Reynolds number $\mathrm{Re}_\tau$, which differs from one flow to another. Consequently, having only $\{\alpha, y^+\}$ in the feature set, the neural network trained in Case 2 can not identify flows at different turbulent Reynolds numbers in the near-wall region where the $y^+$ value is small. 

In Table~\ref{tab:R2results} we  can find a summary of the $R^2$ values for each case study. We can see an obvious improvement of $R^2$ values in Case 2 and Case 4 by adding $y^+$ into the feature set. The difference between Case 2 and Case 4 is not so pronounced in terms of $R^2$ values and especially lies in the test sets, in particular for higher Reynolds number extrapolation test at $\mathrm{Re}_\tau=\np{10000}$. However, we would like to highlight the learning difficulty of Case 2 since we had to push the training epochs till 19,940. This is understandable since the neural network trained in Case 2 was fed with data mixed with different $\mathrm{Re}_\tau$ without explicitly having this information as is in Case 4, and so it could take a long time for the network to reflect this fact. Hence, we conclude that both $y^+$ and $\mathrm{Re}_\tau$ are critical to our model, and we should keep the feature set  $\{\alpha, y^+, \mathrm{Re}_\tau\}$ to yield a better performance in terms of both accuracy and convergence.    

\subsection{$\textbf{T}^{*(0)}$ selection}\label{sec:T0select}
As is mentioned in Section \ref{subsec:TCF}, we question whether there exists an optimal choice of $\textbf{T}^{*(0)}$ for the augmented TBNN architecture. On the basis of former results, we train the augmented TBNN model using different $\textbf{T}^{*(0)}$ with the feature set $\{\alpha, y^+, \mathrm{Re}_\tau\}$ (which yields the best results). The training for Case 5 - Case 8 of Table~\ref{tab:case} is respectively stopped at  9,993, 34,215, 30,990 and 10,943 epochs. 

\begin{figure}[ht]
\centering
\includegraphics[width=1\textwidth]{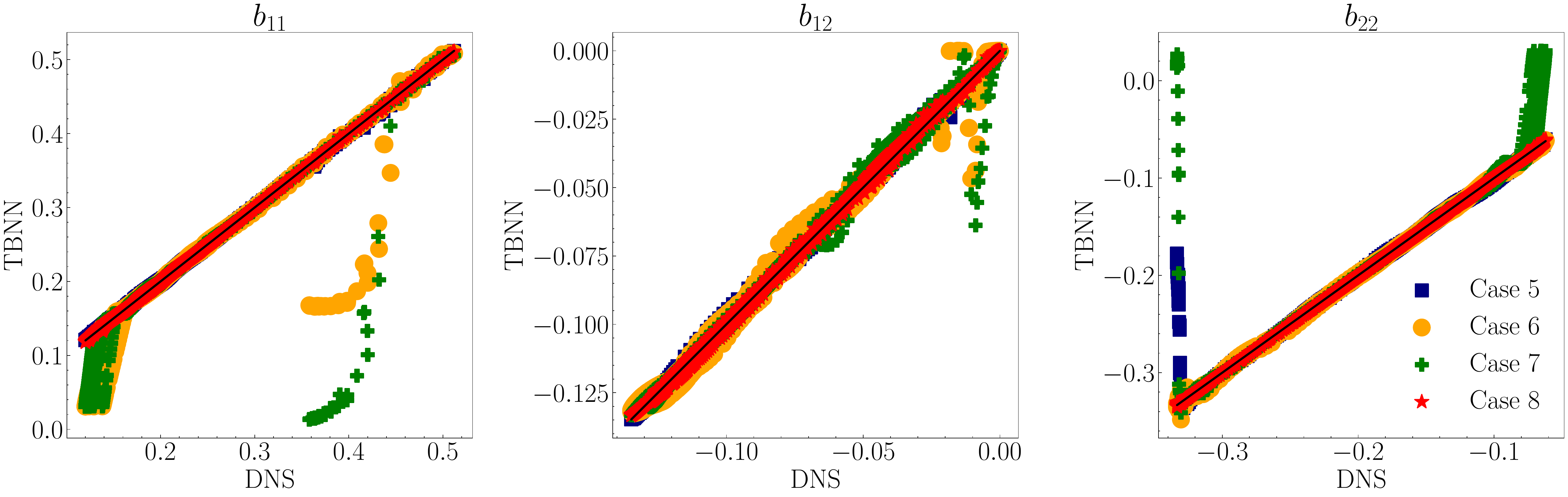}
\caption{$b_{11}$, $b_{12}$ and $b_{22}$ predictions for Case 5 - Case 8 using augmented TBNN models, compared to DNS data, in dispersion plots.}
\label{fig:T0}
\end{figure}

Fig.~\ref{fig:T0} shows the training results for Case 5 - Case 8. Since $b_{33}$ is calculated from $b_{11}$ and $b_{22}$ in the learning process, the corresponding values are not shown. We discover that the performance of the augmented TBNN models is strongly influenced by the $\textbf{T}^{*(0)}$ picked in each case. The models in Cases 5, 6, and 7 with constant $\textbf{T}^{*(0)}$ learn well in general but perform notably poorly in specific value ranges, especially when it comes to diagonal components, $b_{11}$ and $b_{22}$. As can be observed, Case 6 and Case 7 totally fail to predict values near the bounds of $b_{ij}$ intervals, which appears to be a systematic error. By comparing with Fig.~\ref{fig:b_visu_trans}, we identify that those zones refer to either the center of the channel ($y=1$) or the near-wall region. Given that the model trained in Case 5 manages to learn values nearby the channel center, we deduce that the failure at the center for Case 6 and Case 7 is caused by the chosen $\textbf{T}^{*(0)}$ in the model. This is in agreement with our previous analysis based on the Fig.~\ref{fig:b_diag} in Section \ref{subsec:TCF}. Nevertheless, the Case 5 model still fails to predict the $b_{22}$ component in the near-wall region for unclear reasons. We would like to therefore emphasize how much better Case 8 performs by using our newly proposed generalized $\textbf{T}^{*(0)}$, as clearly demonstrated in Fig.~\ref{fig:T0}.

The $R^2$ values for Case 5 - Case 8 can also be found in Table~\ref{tab:R2results}. In accordance with the above observations, the model trained with the generalized $\textbf{T}^{*(0)}$ in Case 8 outperforms those with constant $\textbf{T}^{*(0)}$ values in all aspects. 

\begingroup
\squeezetable
\begin{table*}[ht]
\begin{ruledtabular}
\centering
\caption{\label{tab:R2results}$R^2$ error of $b_{ij}$ predictions for each case study: test 1, 2 and 3 respectively for test set at $\mathrm{Re}_\tau=550; \np{5200}$ and $\np{10000}$.}
\begin{tabular}{lcccccc}
\multicolumn{2}{l}{Case} & $b_{11}$ & $b_{12}$ & $b_{22}$ & $b_{33}$ & Global \\\hline
\multirow{5}{*}{1} & Train & 0.8007 & 0.8755 & 0.7300 & 0.7164 & 0.7807 \\ 
& Test   & 0.8116 & 0.8713 & 0.7340 & 0.5956 & 0.7531\\
& Test 1 & 0.6025 & 0.7808 & 0.3785 & 0.7446 & 0.6266\\
& Test 2 & 0.7990 & 0.8915 & 0.7209 & 0.7854 & 0.7992\\
& Test 3 & 0.8517 & 0.8760 & 0.8062 & 0.4575 & 0.7492\\
\hline
\multirow{5}{*}{2} & Train & 0.9994 & 0.9988 & 0.9996 & 0.9961 & 0.9985 \\ 
& Test & 0.9670 & 0.9837 & 0.9955 & 0.6108 & 0.8892 \\
& Test 1 & 0.9938 & 0.9935 & 0.9985 & 0.9607 & 0.9866 \\ 
& Test 2 & 0.9979 & 0.9990 & 0.9988 & 0.9898 & 0.9964\\
& Test 3 & 0.9440 & 0.9729 & 0.9930 & 0.3255 & 0.8089 \\
\hline
\multirow{5}{*}{3} & Train & 0.8501 & 0.9128 & 0.8027 & 0.7606 & 0.8315 \\ 
& Test & 0.8164 & 0.9095 & 0.7719 & 0.4778 & 0.7439 \\
& Test 1 & 0.6772 & 0.8193 & 0.5494 & 0.7799 & 0.7065 \\
& Test 2 & 0.7945 & 0.9197 & 0.7128 & 0.7963 & 0.8058\\
& Test 3 & 0.8545 & 0.9199 & 0.8770 & 0.2366 & 0.7145\\
\hline
\multirow{5}{*}{4} & \textbf{Train} & \textbf{0.9997} & \textbf{0.9994} & \textbf{0.9999} & \textbf{0.9983} & \textbf{0.9993} \\ 
& \textbf{Test}  & \textbf{0.9937} & \textbf{0.9743} & \textbf{0.9956} & \textbf{0.9646} & \textbf{0.9820} \\
& Test 1 & 0.9904 & 0.9941 & 0.9952 & 0.9753 & 0.9888 \\
& Test 2 & 0.9968 & 0.9983 & 0.9987 & 0.9852 & 0.9947 \\
& Test 3 & 0.9924 & 0.9567 & 0.9939 & 0.9505 & 0.9734 \\
\hline
\multirow{5}{*}{5} & Train & 0.9997 & 0.9989  & 0.9374 & 0.4657 & 0.8504 \\ 
& Test  & 0.9882 & 0.9822 & 0.9405 & 0.4002 & 0.8278 \\
& Test 1 & 0.9919 & 0.9822 & 0.8252 & 0.3217 & 0.7803 \\
& Test 2 & 0.9975 & 0.9967 & 0.9359 & 0.3864 & 0.8291 \\
& Test 3 & 0.9820 & 0.9737 & 0.9641 & 0.4226 & 0.8356 \\
\hline
\multirow{5}{*}{6} & Train & 0.8371 & 0.9939 & 0.9992 & -0.1143 & 0.4217 \\ 
& Test  & 0.8185 & 0.9787 & 0.9171 & -0.2006 & 0.1771 \\
& Test 1 & 0.7483 & 0.9488 & 0.9906 & -0.1101 & 0.3968 \\
& Test 2 & 0.8382 & 0.9963 & 0.9952 & -0.1364 & 0.3665 \\
& Test 3 & 0.8197 & 0.9739 & 0.8580 & -2.5460 & 0.0264 \\
\hline
\multirow{5}{*}{7} & Train & 0.7222 & 0.9910 & 0.5824 & 0.9858 & 0.8204 \\ 
& Test & 0.7428 & 0.9766 & 0.6140 & 0.7561 & 0.7724 \\
& Test 1 & 0.4239 & 0.9595 & -0.0078 & 0.9524 & 0.5820 \\
& Test 2 & 0.7172 & 0.9934 & 0.5618 & 0.9818 & 0.8135 \\
& Test 3 & 0.8158 & 0.9698 & 0.7576 & 0.5885 & 0.7829 \\
\hline
\multirow{5}{*}{8} & \textbf{Train} & \textbf{0.9997} & \textbf{0.9989} & \textbf{0.9999} & \textbf{0.9985} & \textbf{0.9993} \\ 
& \textbf{Test}  & \textbf{0.9911} & \textbf{0.9793} & \textbf{0.9872} & \textbf{0.7843} & \textbf{0.9355} \\
& Test 1 & 0.9930 & 0.9789 & 0.9982 & 0.9720 & 0.9855 \\
& Test 2 & 0.9970 & 0.9976 & 0.9986 & 0.9892 & 0.9956\\
& Test 3 & 0.9873 & 0.9686 & 0.9786 & 0.6304 & 0.8912\\
\end{tabular}
\end{ruledtabular}
\end{table*}
\endgroup

\subsection{Best networks comparison}
\begin{figure}[ht]
\centering
\begin{subfigure}{0.42\textwidth}
    \includegraphics[width=\textwidth]{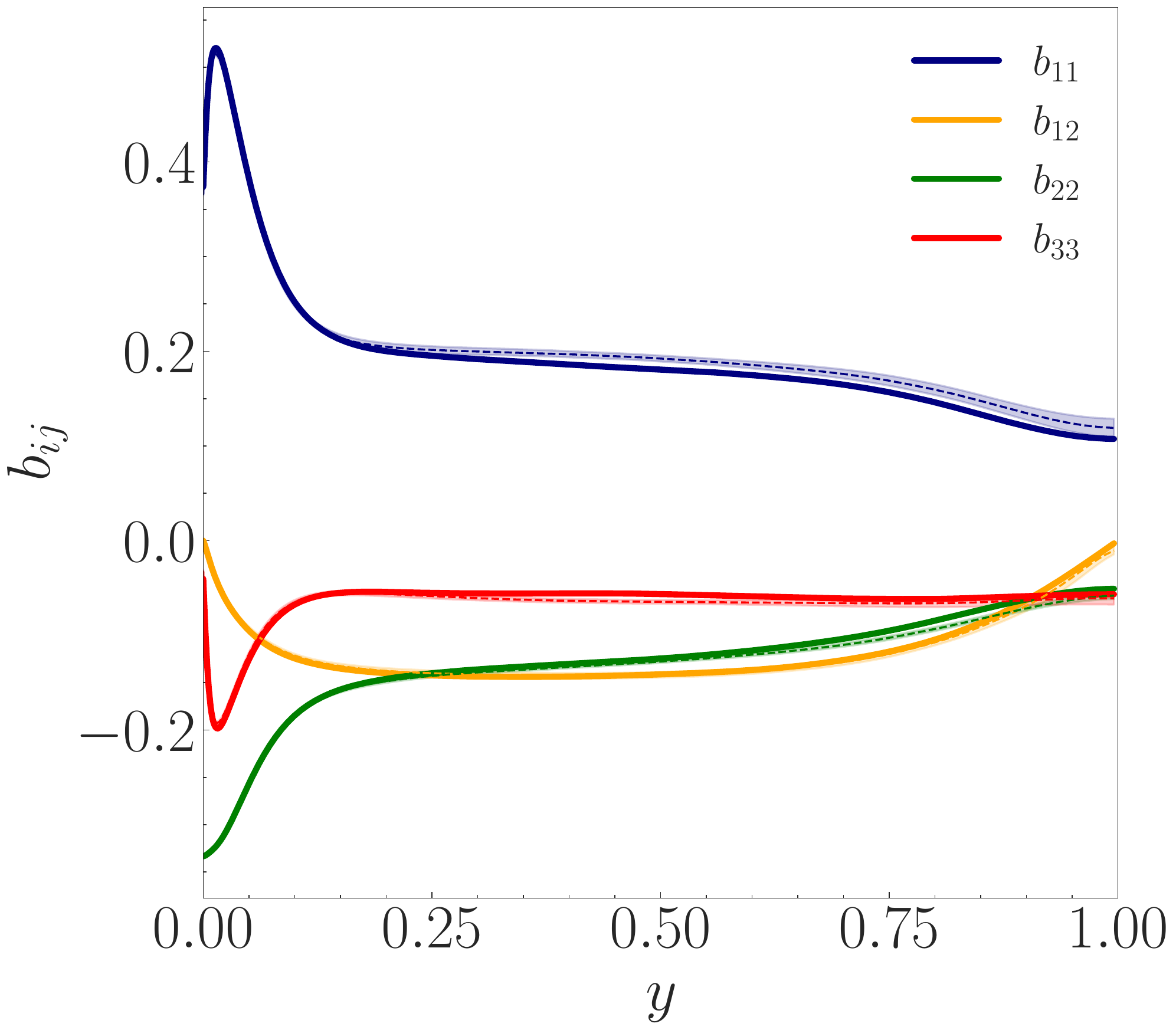}
    \caption{Case 4: Test 1}
    \label{fig:MLP_10runs_T1}
\end{subfigure}
\hfill
\begin{subfigure}{0.42\textwidth}
    \includegraphics[width=\textwidth]{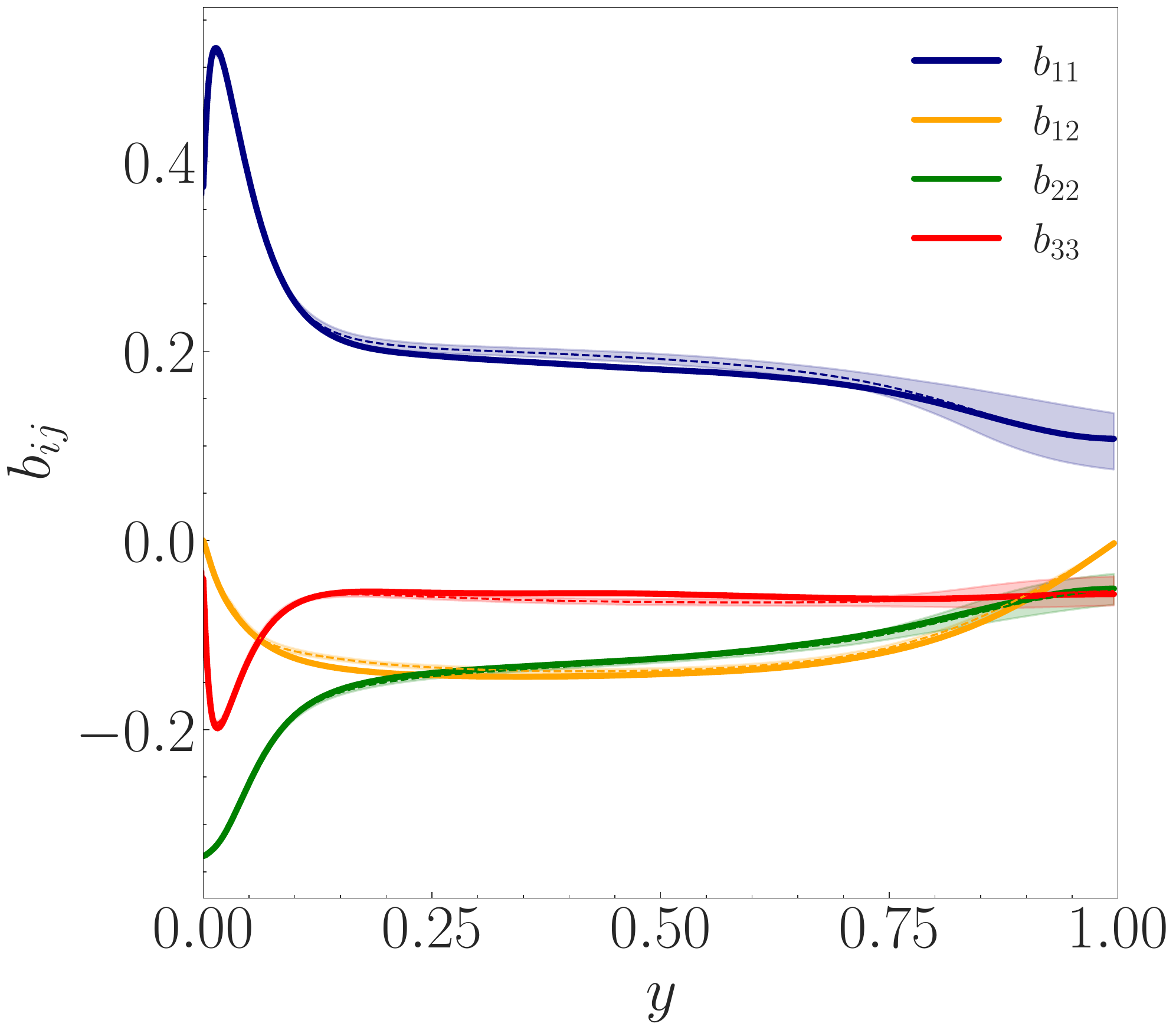}
    \caption{Case 8: Test 1}
    \label{fig:TBNN_10runs_T1}
\end{subfigure}
\hfill
\begin{subfigure}{0.42\textwidth}
    \includegraphics[width=\textwidth]{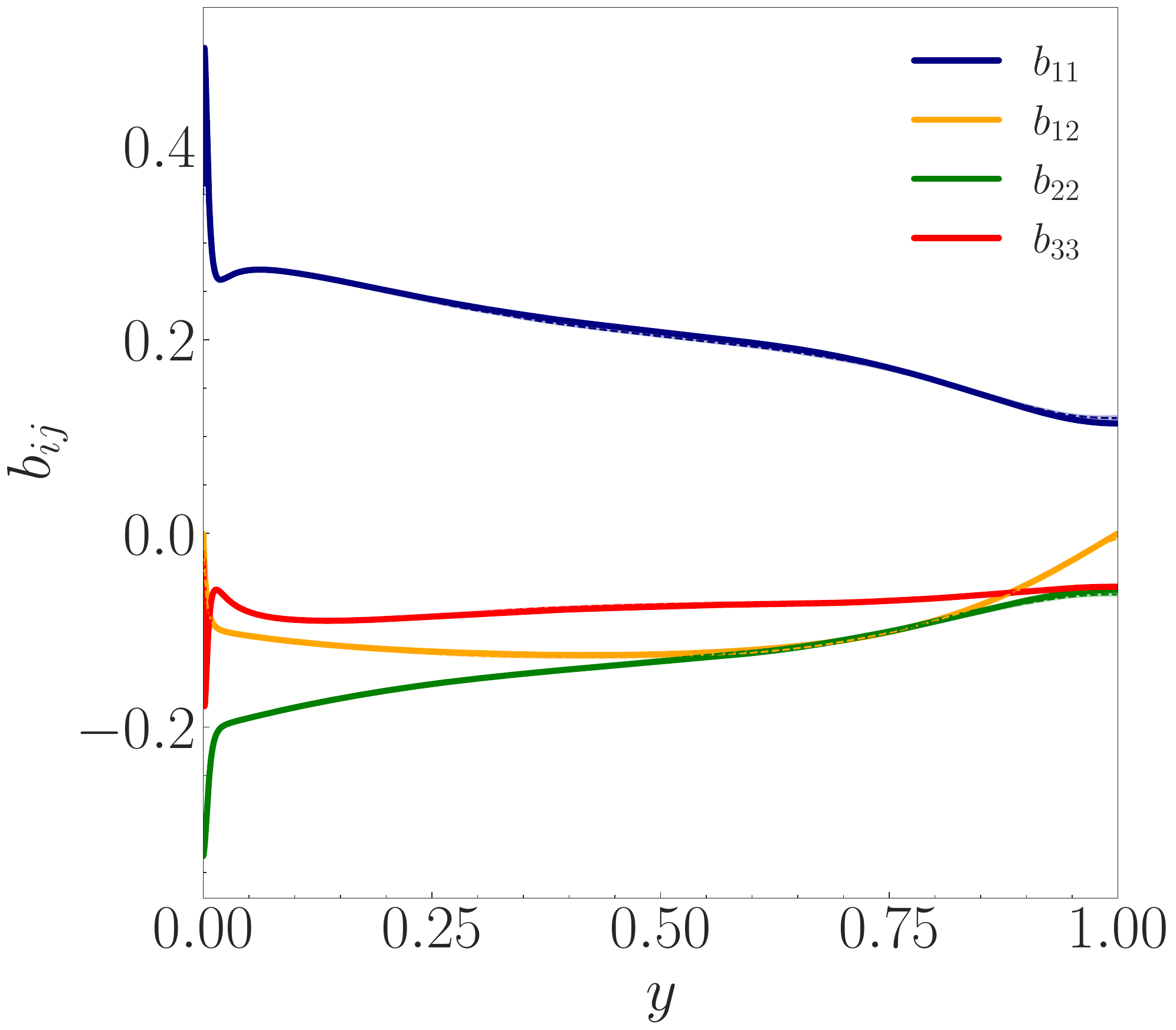}
    \caption{Case 4: Test 2}
    \label{fig:MLP_10runs_T2}
\end{subfigure}
\hfill
\begin{subfigure}{0.42\textwidth}
    \includegraphics[width=\textwidth]{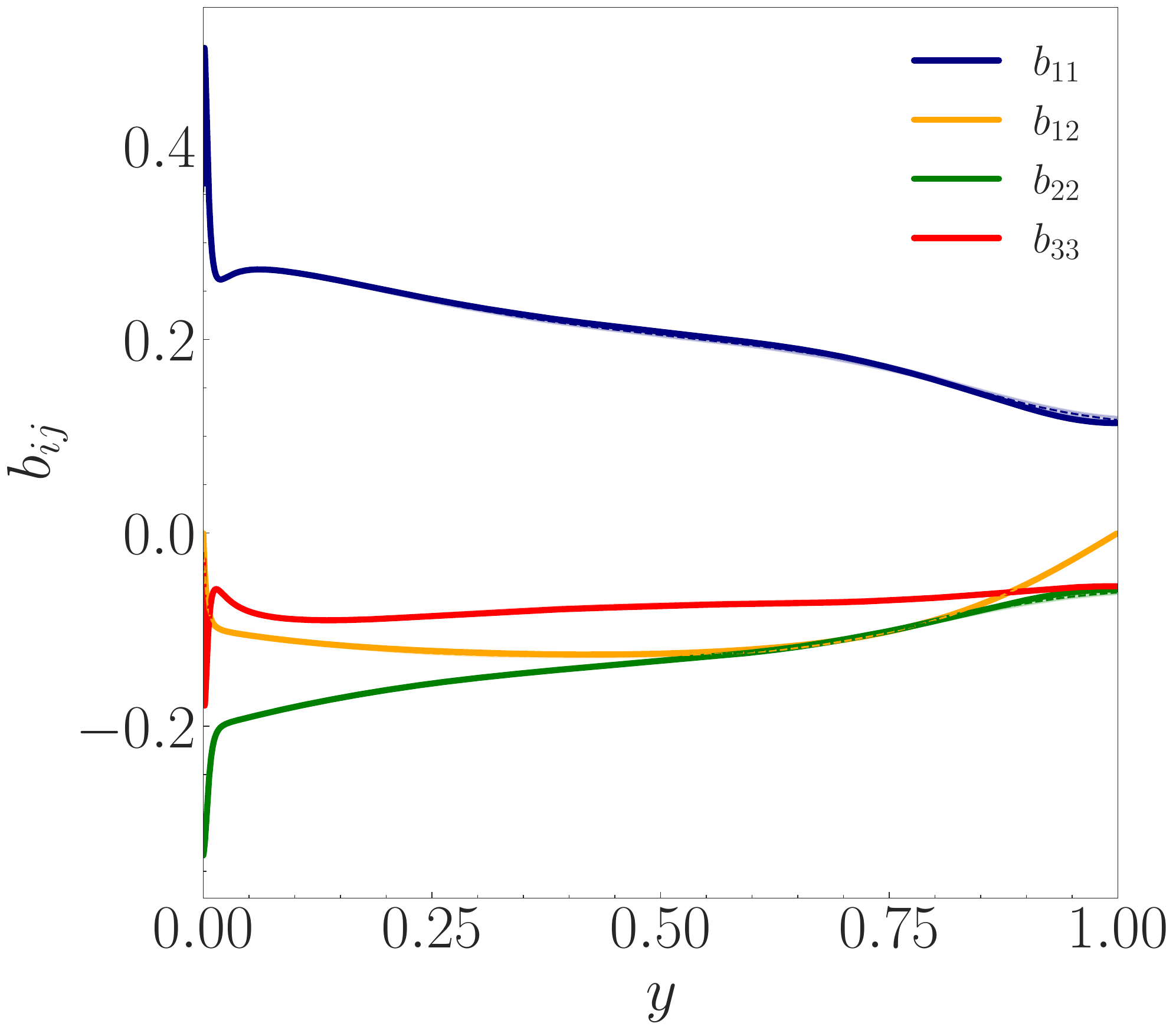}
    \caption{Case 8: Test 2}
    \label{fig:TBNN_10runs_T2}
\end{subfigure}
\hfill
\begin{subfigure}{0.42\textwidth}
    \includegraphics[width=\textwidth]{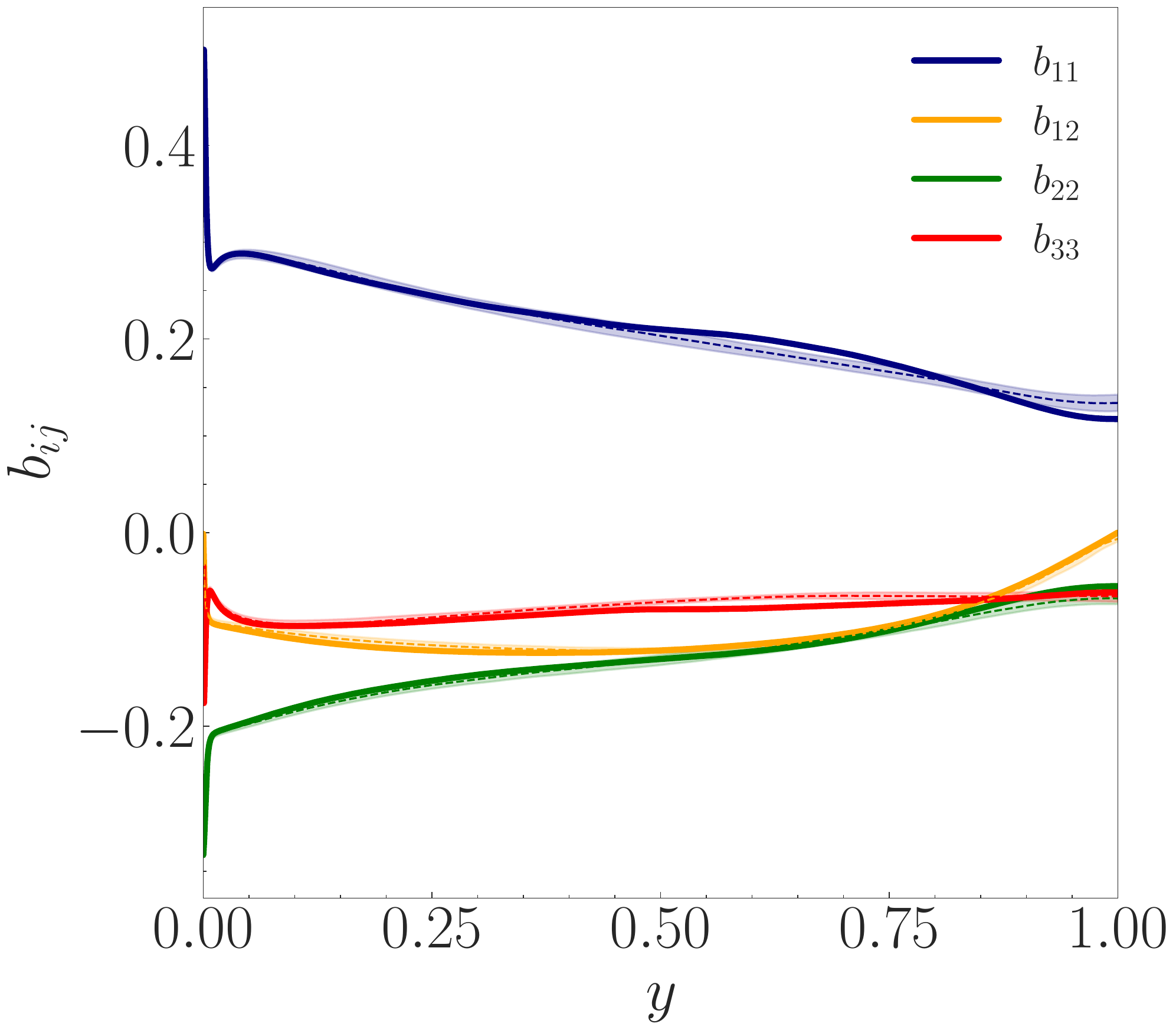}
    \caption{Case 4: Test 3}
    \label{fig:MLP_10runs_T3}
\end{subfigure}
\hfill
\begin{subfigure}{0.42\textwidth}
    \includegraphics[width=\textwidth]{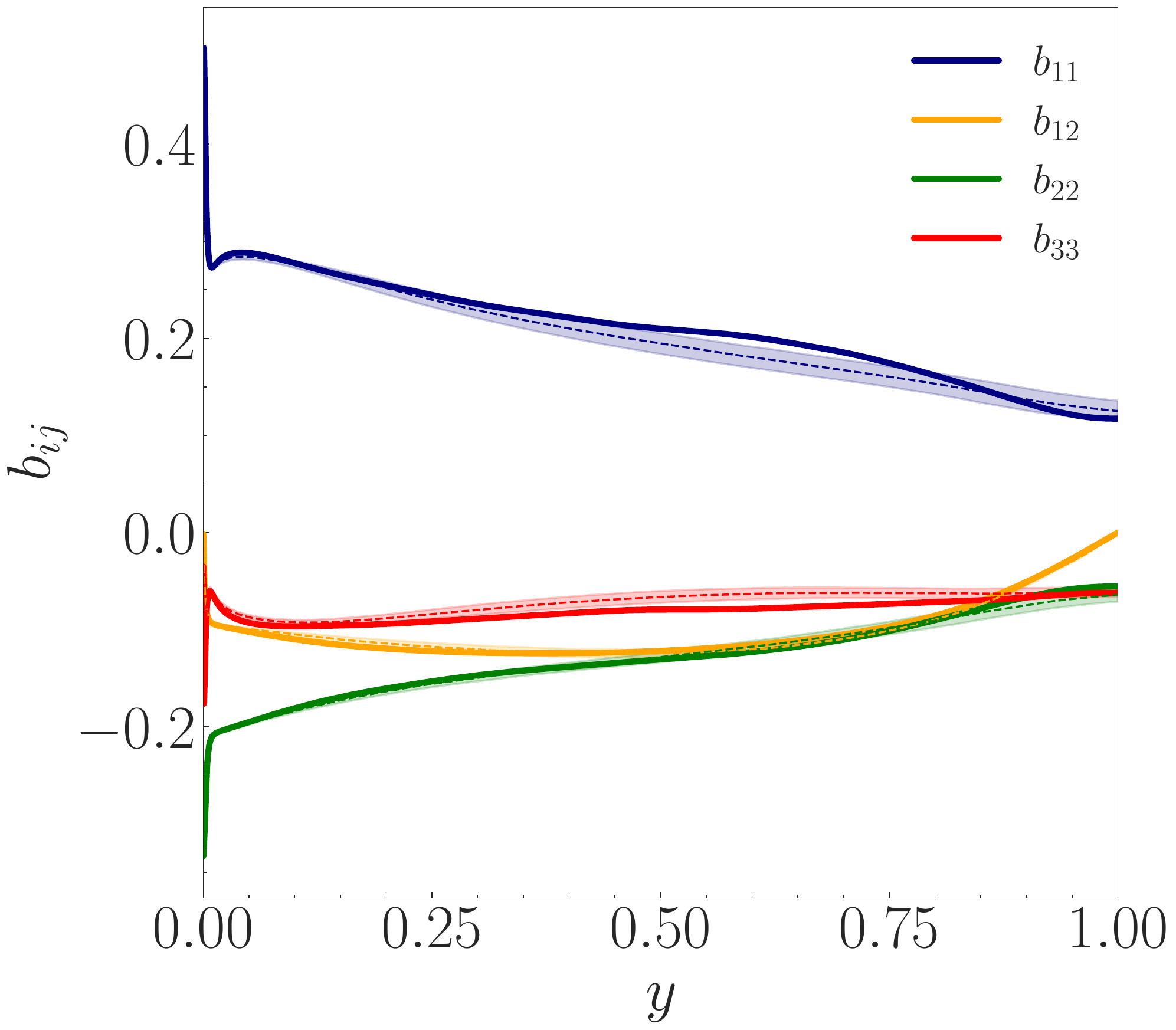}
    \caption{Case 8: Test 3}
    \label{fig:TBNN_10runs_T3}
\end{subfigure}
\hfill        
\caption{Averaged predicted $b_{ij}$ after ten repeated learnings: the DNS data and the averaged predicted values are shown in solid and dotted lines, respectively. The transparent colored region corresponds to the interval of plus and minus one standard deviation.}
\label{fig:10runs}
\end{figure}

The ultimate choice lies between the MLP model trained in Case 4 and the augmented TBNN model trained in Case 8. As shown in Table~\ref{tab:R2results}, both models provide comparable results, except for the prediction of $b_{33}$ component at $\mathrm{Re}_\tau=\np{10000}$. In order to test their robustness, we repeated the learning process ten times for both models. Results are shown in Fig.~\ref{fig:10runs}, in comparison with the DNS data,  where the transparent zones represent the interval of plus and minus one standard deviation to the averaged predictions of these ten learnings.

According to Fig.~\ref{fig:MLP_10runs_T2} and Fig.~\ref{fig:TBNN_10runs_T2}, we can clearly observe that the predicted $b_{ij}$ profiles are nearly identical to the DNS data for the interpolated test, with low standard deviation for both models. Such good predictive performance for the interpolated test at $\mathrm{Re}_\tau=\np{5200}$ has demonstrated the pertinence of our training data, representative enough to make a good prediction for channel flows with an interpolating turbulent Reynolds number. Test results on two extreme extrapolation tests at $\mathrm{Re}_\tau=550$ and $\mathrm{Re}_\tau=\np{10000}$ are shown in Fig.~\ref{fig:MLP_10runs_T1} and Fig.~\ref{fig:MLP_10runs_T3} for Case 4 and in Fig.~\ref{fig:TBNN_10runs_T1} and Fig.~\ref{fig:TBNN_10runs_T3} for Case 8. Even if both models do not perform as well as they do in the interpolating case, the prediction results remain very close and very robust compared to the DNS data only with minor differences. A larger standard deviation of the augmented TBNN model (shown in Fig.~\ref{fig:TBNN_10runs_T1}) is observed near the channel center.

A more quantitative comparison between these two models is shown in Table~\ref{tab:R2Case48}. The averaged performance of our MLP model seems slightly better, especially for the prediction of the $b_{33}$ component at $\mathrm{Re}_\tau=\np{10000}$. Such prediction performance for extrapolating flow configurations is very satisfactory since it is a challenging task to provide extrapolating predictions for flows with varying turbulence levels and characteristics that might be somehow different from the training flows. In addition, these extrapolation results are promising in terms of boosting RANS accuracy for highly turbulent flows, rather than running expensive DNS calculations. 

\begingroup
\squeezetable
\begin{table*}[ht]
\begin{ruledtabular}
\centering
\caption{\label{tab:R2Case48}Averaged $R^2$ error of $b_{ij}$ predictions for Case 4 and Case 8 after ten repeated learnings.}
\begin{tabular}{lcccccc}
\multicolumn{2}{l}{Case} & $b_{11}$ & $b_{12}$ & $b_{22}$ & $b_{33}$ & Global \\\hline
\multirow{5}{*}{4} & Train & 0.9998 & 0.9995 & 0.9999 & 0.9990 & 0.9996 \\ 
& Test  & 0.9857 & 0.9802 & 0.9900 & 0.8612 & 0.9542 \\
& Test 1 & 0.9930 & 0.9931 & 0.9970 & 0.9776 & 0.9902 \\
& Test 2 & 0.9980 & 0.9961 & 0.9990 & 0.9896 & 0.9957 \\
& Test 3 & 0.9771 & 0.9685 & 0.9835 & 0.7649 & 0.9235\\
\hline
\multirow{5}{*}{8} & Train & 0.9998 & 0.9992 & 0.9999 & 0.9989 &  0.9994 \\
& Test   & 0.9690 & 0.9817 & 0.9900 & 0.6770 & 0.9044 \\
& Test 1 & 0.9894 & 0.9832 & 0.9951 & 0.9687 & 0.9841 \\
& Test 2 & 0.9980 & 0.9963 & 0.9990 & 0.9910 & 0.9961 \\
& Test 3 & 0.9483 & 0.9728 & 0.9839 & 0.4403 & 0.8363\\
\end{tabular}
\end{ruledtabular}
\end{table*}
\endgroup

\section{Conclusions}\label{sec:conclusions}
While earlier studies demonstrated the promising possibility of machine-learning-assisted turbulence modeling, some ambiguities still need to be clarified and the prediction of the full anisotropic Reynolds stress tensor remains a challenge, even for the application to the simple turbulent channel flow configuration. The starting point of our study is Ling \textit{et al.}'s Tensor Basis Neural Network (TBNN) \cite{ling2016}, in order to investigate its predictive capability and propose a more adapted neural network, for predicting the full anisotropic Reynolds stress tensor of turbulent channel flow, in both the interpolation and extrapolation scenarios.

For this purpose, we first conducted an input feature selection analysis by utilizing a more flexible neural network architecture of the type of Multi-Layer Perceptron (MLP). We examined its predictive performance with different input feature combinations of the normalized mean velocity gradient $\alpha$, the normalized wall distance $y^+$ and the friction Reynolds number $\mathrm{Re}_\tau$. By cross-comparison, we demonstrated that both $y^+$ and $\mathrm{Re}_\tau$ are essential input features, the former for overcoming the multi-valued problem in the prediction and the latter for discriminating among flows at different $\mathrm{Re}_\tau$ in the near-wall region. 

On the basis of this preliminary analysis, we proposed the augmented TBNN architecture, using the input features early identified. Several ambiguities have been clarified throughout our theoretical and practical investigation on the TBNN. While prior work used the complete version of Pope's general eddy viscosity model \cite{pope1975} for our flow configuration, we observed that the reduced form with one invariant and three tensors is sufficient. We also discovered that TBNN performance is greatly influenced by the constant tensor $\textbf{T}^{*(0)}$, with each of the three alternatives failing more or less either in the near-wall region or in the channel center. According to our finding, only the tensor $\textbf{T}^{*(0)}$ proportional to the diagonal values of the Reynolds stress anisotropy tensor in the channel center is able to provide correct predictions at this location. Even still, the cause for its failure in the near-wall region remains unknown. On the other hand, a considerable enhancement of the augmented TBNN model was achieved by using the generalized $\textbf{T}^{*(0)}$ suggested in this paper. 

Finally, through quantitative comparison between these two model classes, MLP and augmented TBNN, we found that the MLP with $\{\alpha, y^+, \mathrm{Re}_\tau\}$ outperforms the augmented TBNN, which somehow turns out to be a nuanced version of the former, and is more robust. Therefore, we suggest using the MLP proposed in the present work, which provides satisfactory predictions of the full anisotropic Reynolds stress tensor on this particular flow case.

A number of challenges remain for future studies. A straight forward extension of  the present work would be to focus on the hyperparameter tuning in order to consolidate our model. An a-posteriori validation on the resulting flow fields should also be performed, by integrating our neural networks into an in-house developed CFD code \cite{angeli2015, angeli2017}. It would then be interesting to extend our study to more complex flow configurations, including three-dimensionality of the flow statistics and complex phenomena such as recirculations or boundary layer separation and reattachment. To this end, transfer learning or multi-task learning could be explored in future work. 

\section*{Acknowledgements}
This work was granted access to the HPC resources of TGCC under the allocations 2019-A0062A10806 and 2020-A0092A10806 attributed by GENCI (Grand Équipement National de Calcul Intensif).

\bibliography{library}

\begin{thebibliography}{33}%
\makeatletter
\providecommand \@ifxundefined [1]{%
 \@ifx{#1\undefined}
}%
\providecommand \@ifnum [1]{%
 \ifnum #1\expandafter \@firstoftwo
 \else \expandafter \@secondoftwo
 \fi
}%
\providecommand \@ifx [1]{%
 \ifx #1\expandafter \@firstoftwo
 \else \expandafter \@secondoftwo
 \fi
}%
\providecommand \natexlab [1]{#1}%
\providecommand \enquote  [1]{``#1''}%
\providecommand \bibnamefont  [1]{#1}%
\providecommand \bibfnamefont [1]{#1}%
\providecommand \citenamefont [1]{#1}%
\providecommand \href@noop [0]{\@secondoftwo}%
\providecommand \href [0]{\begingroup \@sanitize@url \@href}%
\providecommand \@href[1]{\@@startlink{#1}\@@href}%
\providecommand \@@href[1]{\endgroup#1\@@endlink}%
\providecommand \@sanitize@url [0]{\catcode `\\12\catcode `\$12\catcode
  `\&12\catcode `\#12\catcode `\^12\catcode `\_12\catcode `\%12\relax}%
\providecommand \@@startlink[1]{}%
\providecommand \@@endlink[0]{}%
\providecommand \url  [0]{\begingroup\@sanitize@url \@url }%
\providecommand \@url [1]{\endgroup\@href {#1}{\urlprefix }}%
\providecommand \urlprefix  [0]{URL }%
\providecommand \Eprint [0]{\href }%
\providecommand \doibase [0]{https://doi.org/}%
\providecommand \selectlanguage [0]{\@gobble}%
\providecommand \bibinfo  [0]{\@secondoftwo}%
\providecommand \bibfield  [0]{\@secondoftwo}%
\providecommand \translation [1]{[#1]}%
\providecommand \BibitemOpen [0]{}%
\providecommand \bibitemStop [0]{}%
\providecommand \bibitemNoStop [0]{.\EOS\space}%
\providecommand \EOS [0]{\spacefactor3000\relax}%
\providecommand \BibitemShut  [1]{\csname bibitem#1\endcsname}%
\let\auto@bib@innerbib\@empty
\bibitem [{\citenamefont {Tennekes}\ and\ \citenamefont
  {Lumley}(1972)}]{tennekes1972}%
  \BibitemOpen
  \bibfield  {author} {\bibinfo {author} {\bibfnamefont {H.}~\bibnamefont
  {Tennekes}}\ and\ \bibinfo {author} {\bibfnamefont {J.~L.}\ \bibnamefont
  {Lumley}},\ }\href@noop {} {\emph {\bibinfo {title} {A {{First Course}} in
  {{Turbulence}}}}}\ (\bibinfo  {publisher} {{MIT Press}},\ \bibinfo {address}
  {{Cambridge, MA, USA}},\ \bibinfo {year} {1972})\BibitemShut {NoStop}%
\bibitem [{\citenamefont {Pope}(2000)}]{pope2000}%
  \BibitemOpen
  \bibfield  {author} {\bibinfo {author} {\bibfnamefont {S.~B.}\ \bibnamefont
  {Pope}},\ }\href@noop {} {\emph {\bibinfo {title} {Turbulent {{Flows}}}}}\
  (\bibinfo  {publisher} {{Cambridge University Press}},\ \bibinfo {year}
  {2000})\BibitemShut {NoStop}%
\bibitem [{\citenamefont {Kronborg}\ \emph {et~al.}(2022)\citenamefont
  {Kronborg}, \citenamefont {Svelander}, \citenamefont {Eriksson-Lidbrink},
  \citenamefont {Lindström}, \citenamefont {Homs-Pons}, \citenamefont
  {Lucor},\ and\ \citenamefont {Hoffman}}]{kronborg2022}%
  \BibitemOpen
  \bibfield  {author} {\bibinfo {author} {\bibfnamefont {J.}~\bibnamefont
  {Kronborg}}, \bibinfo {author} {\bibfnamefont {F.}~\bibnamefont {Svelander}},
  \bibinfo {author} {\bibfnamefont {S.}~\bibnamefont {Eriksson-Lidbrink}},
  \bibinfo {author} {\bibfnamefont {L.}~\bibnamefont {Lindström}}, \bibinfo
  {author} {\bibfnamefont {C.}~\bibnamefont {Homs-Pons}}, \bibinfo {author}
  {\bibfnamefont {D.}~\bibnamefont {Lucor}},\ and\ \bibinfo {author}
  {\bibfnamefont {J.}~\bibnamefont {Hoffman}},\ }\bibfield  {title} {\bibinfo
  {title} {Computational {{Analysis}} of {{Flow Structures}} in {{Turbulent
  Ventricular Blood Flow Associated With Mitral Valve Intervention}}},\
  }\href@noop {} {\bibfield  {journal} {\bibinfo  {journal} {Front. Physiol.}\
  }\textbf {\bibinfo {volume} {13}} (\bibinfo {year} {2022})}\BibitemShut
  {NoStop}%
\bibitem [{int(2022)}]{international2022iaea}%
  \BibitemOpen
  \href@noop {} {\emph {\bibinfo {title} {Summary {{Review}} on the
  {{Application}} of {{Computational Fluid Dynamics}} in {{Nuclear Power Plant
  Design}}}}},\ \bibinfo {series} {Nuclear Energy Series}\ No.\ \bibinfo
  {number} {NR-T-1.20}\ (\bibinfo  {publisher} {{International Atomic Energy
  Agency}},\ \bibinfo {address} {{Vienna}},\ \bibinfo {year}
  {2022})\BibitemShut {NoStop}%
\bibitem [{\citenamefont {Chapman}(1979)}]{chapman1979}%
  \BibitemOpen
  \bibfield  {author} {\bibinfo {author} {\bibfnamefont {D.~R.}\ \bibnamefont
  {Chapman}},\ }\bibfield  {title} {\bibinfo {title} {Computational
  {{Aerodynamics Development}} and {{Outlook}}},\ }\href@noop {} {\bibfield
  {journal} {\bibinfo  {journal} {AIAA J.}\ }\textbf {\bibinfo {volume} {17}},\
  \bibinfo {pages} {1293} (\bibinfo {year} {1979})}\BibitemShut {NoStop}%
\bibitem [{\citenamefont {Johansson}(2002)}]{johansson2002}%
  \BibitemOpen
  \bibfield  {author} {\bibinfo {author} {\bibfnamefont {A.}~\bibnamefont
  {Johansson}},\ }\bibfield  {title} {\bibinfo {title} {Engineering
  {{Turbulence Models}} and their {{Development}}, with {{Emphasis}} on
  {{Explicit Algebraic Reynolds Stress Models}}},\ }in\ \href@noop {} {\emph
  {\bibinfo {booktitle} {Theories of {{Turbulence}}}}},\ \bibinfo {series and
  number} {International {{Centre}} for {{Mechanical Sciences}}},\ \bibinfo
  {editor} {edited by\ \bibinfo {editor} {\bibfnamefont {M.}~\bibnamefont
  {Oberlack}}\ and\ \bibinfo {editor} {\bibfnamefont {F.~H.}\ \bibnamefont
  {Busse}}}\ (\bibinfo  {publisher} {{Springer}},\ \bibinfo {address}
  {{Vienna}},\ \bibinfo {year} {2002})\ pp.\ \bibinfo {pages}
  {253--300}\BibitemShut {NoStop}%
\bibitem [{\citenamefont {Launder}\ and\ \citenamefont
  {Spalding}(1974)}]{launder1974}%
  \BibitemOpen
  \bibfield  {author} {\bibinfo {author} {\bibfnamefont {B.}~\bibnamefont
  {Launder}}\ and\ \bibinfo {author} {\bibfnamefont {D.}~\bibnamefont
  {Spalding}},\ }\bibfield  {title} {\bibinfo {title} {The {{Numerical
  Computation}} of {{Turbulent Flows}}},\ }\href@noop {} {\bibfield  {journal}
  {\bibinfo  {journal} {Comput. Methods Appl. Mech. Eng.}\ }\textbf {\bibinfo
  {volume} {3}},\ \bibinfo {pages} {269} (\bibinfo {year} {1974})}\BibitemShut
  {NoStop}%
\bibitem [{\citenamefont {Duraisamy}\ \emph {et~al.}(1 05)\citenamefont
  {Duraisamy}, \citenamefont {Iaccarino},\ and\ \citenamefont
  {Xiao}}]{duraisamy2019}%
  \BibitemOpen
  \bibfield  {author} {\bibinfo {author} {\bibfnamefont {K.}~\bibnamefont
  {Duraisamy}}, \bibinfo {author} {\bibfnamefont {G.}~\bibnamefont
  {Iaccarino}},\ and\ \bibinfo {author} {\bibfnamefont {H.}~\bibnamefont
  {Xiao}},\ }\bibfield  {title} {\bibinfo {title} {Turbulence {{Modeling}} in
  the {{Age}} of {{Data}}},\ }\href@noop {} {\bibfield  {journal} {\bibinfo
  {journal} {Annu. Rev. Fluid Mech.}\ }\textbf {\bibinfo {volume} {51}},\
  \bibinfo {pages} {357} (\bibinfo {year} {2019-01-05})}\BibitemShut {NoStop}%
\bibitem [{\citenamefont {Brenner}\ \emph {et~al.}(2019)\citenamefont
  {Brenner}, \citenamefont {Eldredge},\ and\ \citenamefont
  {Freund}}]{brenner2019}%
  \BibitemOpen
  \bibfield  {author} {\bibinfo {author} {\bibfnamefont {M.~P.}\ \bibnamefont
  {Brenner}}, \bibinfo {author} {\bibfnamefont {J.~D.}\ \bibnamefont
  {Eldredge}},\ and\ \bibinfo {author} {\bibfnamefont {J.~B.}\ \bibnamefont
  {Freund}},\ }\bibfield  {title} {\bibinfo {title} {Perspective on {{Machine
  Learning}} for {{Advancing Fluid Mechanics}}},\ }\href@noop {} {\bibfield
  {journal} {\bibinfo  {journal} {Phys. Rev. Fluids}\ }\textbf {\bibinfo
  {volume} {4}},\ \bibinfo {pages} {100501} (\bibinfo {year}
  {2019})}\BibitemShut {NoStop}%
\bibitem [{\citenamefont {Brunton}\ \emph {et~al.}(2020)\citenamefont
  {Brunton}, \citenamefont {Noack},\ and\ \citenamefont
  {Koumoutsakos}}]{brunton2020}%
  \BibitemOpen
  \bibfield  {author} {\bibinfo {author} {\bibfnamefont {S.~L.}\ \bibnamefont
  {Brunton}}, \bibinfo {author} {\bibfnamefont {B.~R.}\ \bibnamefont {Noack}},\
  and\ \bibinfo {author} {\bibfnamefont {P.}~\bibnamefont {Koumoutsakos}},\
  }\bibfield  {title} {\bibinfo {title} {Machine {{Learning}} for {{Fluid
  Mechanics}}},\ }\href@noop {} {\bibfield  {journal} {\bibinfo  {journal}
  {Annu. Rev. Fluid Mech.}\ }\textbf {\bibinfo {volume} {52}},\ \bibinfo
  {pages} {477} (\bibinfo {year} {2020})}\BibitemShut {NoStop}%
\bibitem [{\citenamefont {Duraisamy}(2021)}]{duraisamy2021}%
  \BibitemOpen
  \bibfield  {author} {\bibinfo {author} {\bibfnamefont {K.}~\bibnamefont
  {Duraisamy}},\ }\bibfield  {title} {\bibinfo {title} {Perspectives on
  {{Machine Learning-Augmented Reynolds-Averaged}} and {{Large Eddy Simulation
  Models}} of {{Turbulence}}},\ }\href@noop {} {\bibfield  {journal} {\bibinfo
  {journal} {Phys. Rev. Fluids}\ }\textbf {\bibinfo {volume} {6}},\ \bibinfo
  {pages} {050504} (\bibinfo {year} {2021})}\BibitemShut {NoStop}%
\bibitem [{\citenamefont {Ling}\ \emph {et~al.}(201)\citenamefont {Ling},
  \citenamefont {Kurzawski},\ and\ \citenamefont {Templeton}}]{ling2016}%
  \BibitemOpen
  \bibfield  {author} {\bibinfo {author} {\bibfnamefont {J.}~\bibnamefont
  {Ling}}, \bibinfo {author} {\bibfnamefont {A.}~\bibnamefont {Kurzawski}},\
  and\ \bibinfo {author} {\bibfnamefont {J.}~\bibnamefont {Templeton}},\
  }\bibfield  {title} {\bibinfo {title} {Reynolds {{Averaged Turbulence
  Modelling Using Deep Neural Networks}} with {{Embedded Invariance}}},\
  }\href@noop {} {\bibfield  {journal} {\bibinfo  {journal} {J. Fluid Mech.}\
  }\textbf {\bibinfo {volume} {807}},\ \bibinfo {pages} {155} (\bibinfo {year}
  {201})}\BibitemShut {NoStop}%
\bibitem [{\citenamefont {Pope}(1975)}]{pope1975}%
  \BibitemOpen
  \bibfield  {author} {\bibinfo {author} {\bibfnamefont {S.~B.}\ \bibnamefont
  {Pope}},\ }\bibfield  {title} {\bibinfo {title} {A {{More General
  Effective-Viscosity Hypothesis}}},\ }\href@noop {} {\bibfield  {journal}
  {\bibinfo  {journal} {J. Fluid Mech.}\ }\textbf {\bibinfo {volume} {72}},\
  \bibinfo {pages} {331} (\bibinfo {year} {1975})}\BibitemShut {NoStop}%
\bibitem [{\citenamefont {Fang}\ \emph {et~al.}(2020)\citenamefont {Fang},
  \citenamefont {Sondak}, \citenamefont {Protopapas},\ and\ \citenamefont
  {Succi}}]{fang2020}%
  \BibitemOpen
  \bibfield  {author} {\bibinfo {author} {\bibfnamefont {R.}~\bibnamefont
  {Fang}}, \bibinfo {author} {\bibfnamefont {D.}~\bibnamefont {Sondak}},
  \bibinfo {author} {\bibfnamefont {P.}~\bibnamefont {Protopapas}},\ and\
  \bibinfo {author} {\bibfnamefont {S.}~\bibnamefont {Succi}},\ }\bibfield
  {title} {\bibinfo {title} {Neural {{Network Models}} for the {{Anisotropic
  Reynolds Stress Tensor}} in {{Turbulent Channel Flow}}},\ }\href@noop {}
  {\bibfield  {journal} {\bibinfo  {journal} {J. Turbul.}\ }\textbf {\bibinfo
  {volume} {21}},\ \bibinfo {pages} {525} (\bibinfo {year} {2020})}\BibitemShut
  {NoStop}%
\bibitem [{\citenamefont {Sáez~de Ocáriz~Borde}\ \emph {et~al.}(1
  10)\citenamefont {Sáez~de Ocáriz~Borde}, \citenamefont {Sondak},\ and\
  \citenamefont {Protopapas}}]{saezdeocarizborde2021}%
  \BibitemOpen
  \bibfield  {author} {\bibinfo {author} {\bibfnamefont {H.}~\bibnamefont
  {Sáez~de Ocáriz~Borde}}, \bibinfo {author} {\bibfnamefont {D.}~\bibnamefont
  {Sondak}},\ and\ \bibinfo {author} {\bibfnamefont {P.}~\bibnamefont
  {Protopapas}},\ }\bibfield  {title} {\bibinfo {title} {Convolutional {{Neural
  Network Models}} and {{Interpretability}} for the {{Anisotropic Reynolds
  Stress Tensor}} in {{Turbulent One-Dimensional Flows}}},\ }\href@noop {}
  {\bibfield  {journal} {\bibinfo  {journal} {J. Turbul.}\ }\textbf {\bibinfo
  {volume} {23}},\ \bibinfo {pages} {1} (\bibinfo {year}
  {2021-11-10})}\BibitemShut {NoStop}%
\bibitem [{\citenamefont {Sáez~de Ocáriz~Borde}\ \emph
  {et~al.}(2022)\citenamefont {Sáez~de Ocáriz~Borde}, \citenamefont
  {Sondak},\ and\ \citenamefont {Protopapas}}]{saezdeocarizborde2022}%
  \BibitemOpen
  \bibfield  {author} {\bibinfo {author} {\bibfnamefont {H.}~\bibnamefont
  {Sáez~de Ocáriz~Borde}}, \bibinfo {author} {\bibfnamefont {D.}~\bibnamefont
  {Sondak}},\ and\ \bibinfo {author} {\bibfnamefont {P.}~\bibnamefont
  {Protopapas}},\ }\bibfield  {title} {\bibinfo {title} {Multi-{{Task
  Learning}} based {{Convolutional Models}} with {{Curriculum Learning}} for
  the {{Anisotropic Reynolds Stress Tensor}} in {{Turbulent Duct Flow}}},\
  }\Eprint {https://arxiv.org/abs/2111.00328} {arXiv:2111.00328}  (\bibinfo
  {year} {2022})\BibitemShut {NoStop}%
\bibitem [{\citenamefont {Jiang}\ \emph {et~al.}(202)\citenamefont {Jiang},
  \citenamefont {Vinuesa}, \citenamefont {Chen}, \citenamefont {Mi},
  \citenamefont {Laima},\ and\ \citenamefont {Li}}]{jiang2021}%
  \BibitemOpen
  \bibfield  {author} {\bibinfo {author} {\bibfnamefont {C.}~\bibnamefont
  {Jiang}}, \bibinfo {author} {\bibfnamefont {R.}~\bibnamefont {Vinuesa}},
  \bibinfo {author} {\bibfnamefont {R.}~\bibnamefont {Chen}}, \bibinfo {author}
  {\bibfnamefont {J.}~\bibnamefont {Mi}}, \bibinfo {author} {\bibfnamefont
  {S.}~\bibnamefont {Laima}},\ and\ \bibinfo {author} {\bibfnamefont
  {H.}~\bibnamefont {Li}},\ }\bibfield  {title} {\bibinfo {title} {An
  {{Interpretable Framework}} of {{Data-Driven Turbulence Modeling Using Deep
  Neural Networks}}},\ }\href@noop {} {\bibfield  {journal} {\bibinfo
  {journal} {Phys. Fluids}\ }\textbf {\bibinfo {volume} {33}},\ \bibinfo
  {pages} {055133} (\bibinfo {year} {202})}\BibitemShut {NoStop}%
\bibitem [{\citenamefont {Emory}\ \emph {et~al.}(2013)\citenamefont {Emory},
  \citenamefont {Larsson},\ and\ \citenamefont {Iaccarino}}]{emory2013}%
  \BibitemOpen
  \bibfield  {author} {\bibinfo {author} {\bibfnamefont {M.}~\bibnamefont
  {Emory}}, \bibinfo {author} {\bibfnamefont {J.}~\bibnamefont {Larsson}},\
  and\ \bibinfo {author} {\bibfnamefont {G.}~\bibnamefont {Iaccarino}},\
  }\bibfield  {title} {\bibinfo {title} {Modeling of {{Structural
  Uncertainties}} in {{Reynolds-Averaged Navier-Stokes Closures}}},\
  }\href@noop {} {\bibfield  {journal} {\bibinfo  {journal} {Phys. Fluids}\
  }\textbf {\bibinfo {volume} {25}},\ \bibinfo {pages} {110822} (\bibinfo
  {year} {2013})}\BibitemShut {NoStop}%
\bibitem [{\citenamefont {Xiao}\ and\ \citenamefont
  {Cinnella}(2019)}]{xiao2019}%
  \BibitemOpen
  \bibfield  {author} {\bibinfo {author} {\bibfnamefont {H.}~\bibnamefont
  {Xiao}}\ and\ \bibinfo {author} {\bibfnamefont {P.}~\bibnamefont
  {Cinnella}},\ }\bibfield  {title} {\bibinfo {title} {Quantification of
  {{Model Uncertainty}} in {{RANS Simulations}}: {{A Review}}},\ }\href@noop {}
  {\bibfield  {journal} {\bibinfo  {journal} {Prog. Aerosp. Sci.}\ }\textbf
  {\bibinfo {volume} {108}},\ \bibinfo {pages} {1} (\bibinfo {year}
  {2019})}\BibitemShut {NoStop}%
\bibitem [{\citenamefont {Wang}\ \emph {et~al.}(2018)\citenamefont {Wang},
  \citenamefont {Wu}, \citenamefont {Ling}, \citenamefont {Iaccarino},\ and\
  \citenamefont {Xiao}}]{wang2018}%
  \BibitemOpen
  \bibfield  {author} {\bibinfo {author} {\bibfnamefont {J.}~\bibnamefont
  {Wang}}, \bibinfo {author} {\bibfnamefont {J.}~\bibnamefont {Wu}}, \bibinfo
  {author} {\bibfnamefont {J.}~\bibnamefont {Ling}}, \bibinfo {author}
  {\bibfnamefont {G.}~\bibnamefont {Iaccarino}},\ and\ \bibinfo {author}
  {\bibfnamefont {H.}~\bibnamefont {Xiao}},\ }\bibfield  {title} {\bibinfo
  {title} {A {{Comprehensive Physics-Informed Machine Learning Framework}} for
  {{Predictive Turbulence Modeling}}},\ }\href@noop {} {\bibfield  {journal}
  {\bibinfo  {journal} {Phys. Rev. Fluids}\ }\textbf {\bibinfo {volume} {3}},\
  \bibinfo {pages} {074602} (\bibinfo {year} {2018})}\BibitemShut {NoStop}%
\bibitem [{\citenamefont {Wu}\ \emph {et~al.}(2019)\citenamefont {Wu},
  \citenamefont {Xiao}, \citenamefont {Sun},\ and\ \citenamefont
  {Wang}}]{wu2019b}%
  \BibitemOpen
  \bibfield  {author} {\bibinfo {author} {\bibfnamefont {J.}~\bibnamefont
  {Wu}}, \bibinfo {author} {\bibfnamefont {H.}~\bibnamefont {Xiao}}, \bibinfo
  {author} {\bibfnamefont {R.}~\bibnamefont {Sun}},\ and\ \bibinfo {author}
  {\bibfnamefont {Q.}~\bibnamefont {Wang}},\ }\bibfield  {title} {\bibinfo
  {title} {Reynolds-averaged {{Navier}}–{{Stokes}} equations with explicit
  data-driven {{Reynolds}} stress closure can be ill-conditioned},\ }\href@noop
  {} {\bibfield  {journal} {\bibinfo  {journal} {J. Fluid Mech.}\ }\textbf
  {\bibinfo {volume} {869}},\ \bibinfo {pages} {553} (\bibinfo {year}
  {2019})}\BibitemShut {NoStop}%
\bibitem [{\citenamefont {Jin}\ \emph {et~al.}(2021)\citenamefont {Jin},
  \citenamefont {Cai}, \citenamefont {Li},\ and\ \citenamefont
  {Karniadakis}}]{jin2021}%
  \BibitemOpen
  \bibfield  {author} {\bibinfo {author} {\bibfnamefont {X.}~\bibnamefont
  {Jin}}, \bibinfo {author} {\bibfnamefont {S.}~\bibnamefont {Cai}}, \bibinfo
  {author} {\bibfnamefont {H.}~\bibnamefont {Li}},\ and\ \bibinfo {author}
  {\bibfnamefont {G.~E.}\ \bibnamefont {Karniadakis}},\ }\bibfield  {title}
  {\bibinfo {title} {Nsfnets ({{Navier-Stokes}} flow nets): {{Physics-Informed
  Neural Networks}} for the {{Incompressible Navier-Stokes Equations}}},\
  }\href@noop {} {\bibfield  {journal} {\bibinfo  {journal} {J. Comput. Phys.}\
  }\textbf {\bibinfo {volume} {426}},\ \bibinfo {pages} {109951} (\bibinfo
  {year} {2021})}\BibitemShut {NoStop}%
\bibitem [{\citenamefont {Lucor}\ \emph {et~al.}(2022)\citenamefont {Lucor},
  \citenamefont {Agrawal},\ and\ \citenamefont {Sergent}}]{lucor2022}%
  \BibitemOpen
  \bibfield  {author} {\bibinfo {author} {\bibfnamefont {D.}~\bibnamefont
  {Lucor}}, \bibinfo {author} {\bibfnamefont {A.}~\bibnamefont {Agrawal}},\
  and\ \bibinfo {author} {\bibfnamefont {A.}~\bibnamefont {Sergent}},\
  }\bibfield  {title} {\bibinfo {title} {Simple {{Computational Strategies}}
  for {{More Effective Physics-Informed Neural Networks Modeling}} of
  {{Turbulent Natural Convection}}},\ }\href@noop {} {\bibfield  {journal}
  {\bibinfo  {journal} {J. Comput. Phys.}\ }\textbf {\bibinfo {volume} {456}},\
  \bibinfo {pages} {111022} (\bibinfo {year} {2022})}\BibitemShut {NoStop}%
\bibitem [{\citenamefont {Boussinesq}(1897)}]{boussinesq1897a}%
  \BibitemOpen
  \bibfield  {author} {\bibinfo {author} {\bibfnamefont {J.}~\bibnamefont
  {Boussinesq}},\ }\href@noop {} {\bibinfo {title} {Théorie de l'{{Écoulement
  Tourbillonnant}} et {{Tumultueux}} des {{Liquides}} dans les {{Lits
  Rectilignes}} à {{Grande Section}}}} (\bibinfo {year} {1897})\BibitemShut
  {NoStop}%
\bibitem [{\citenamefont {Craft}\ \emph {et~al.}(1996)\citenamefont {Craft},
  \citenamefont {Launder},\ and\ \citenamefont {Suga}}]{craft1996}%
  \BibitemOpen
  \bibfield  {author} {\bibinfo {author} {\bibfnamefont {T.}~\bibnamefont
  {Craft}}, \bibinfo {author} {\bibfnamefont {B.}~\bibnamefont {Launder}},\
  and\ \bibinfo {author} {\bibfnamefont {K.}~\bibnamefont {Suga}},\ }\bibfield
  {title} {\bibinfo {title} {Development and {{Application}} of a {{Cubic
  Eddy-Viscosity Model}} of {{Turbulence}}},\ }\href@noop {} {\bibfield
  {journal} {\bibinfo  {journal} {Int. J. Heat Fluid Flow}\ }\textbf {\bibinfo
  {volume} {17}},\ \bibinfo {pages} {108} (\bibinfo {year} {1996})}\BibitemShut
  {NoStop}%
\bibitem [{\citenamefont {Liu}\ \emph {et~al.}(2021)\citenamefont {Liu},
  \citenamefont {Fang}, \citenamefont {Rolfo}, \citenamefont {Moulinec},\ and\
  \citenamefont {Emerson}}]{liu2021a}%
  \BibitemOpen
  \bibfield  {author} {\bibinfo {author} {\bibfnamefont {W.}~\bibnamefont
  {Liu}}, \bibinfo {author} {\bibfnamefont {J.}~\bibnamefont {Fang}}, \bibinfo
  {author} {\bibfnamefont {S.}~\bibnamefont {Rolfo}}, \bibinfo {author}
  {\bibfnamefont {C.}~\bibnamefont {Moulinec}},\ and\ \bibinfo {author}
  {\bibfnamefont {D.~R.}\ \bibnamefont {Emerson}},\ }\bibfield  {title}
  {\bibinfo {title} {An {{Iterative Machine-Learning Framework}} for {{RANS
  Turbulence Modeling}}},\ }\href@noop {} {\bibfield  {journal} {\bibinfo
  {journal} {Int. J. Heat Fluid Flow}\ }\textbf {\bibinfo {volume} {90}},\
  \bibinfo {pages} {108822} (\bibinfo {year} {2021})}\BibitemShut {NoStop}%
\bibitem [{\citenamefont {Moser}\ \emph {et~al.}(1999)\citenamefont {Moser},
  \citenamefont {Kim},\ and\ \citenamefont {Mansour}}]{moser1999}%
  \BibitemOpen
  \bibfield  {author} {\bibinfo {author} {\bibfnamefont {R.~D.}\ \bibnamefont
  {Moser}}, \bibinfo {author} {\bibfnamefont {J.}~\bibnamefont {Kim}},\ and\
  \bibinfo {author} {\bibfnamefont {N.~N.}\ \bibnamefont {Mansour}},\
  }\bibfield  {title} {\bibinfo {title} {Direct {{Numerical Simulation}} of
  {{Turbulent Channel Flow}} up to {$Re_\tau \approx 5200$}},\ }\href@noop {}
  {\bibfield  {journal} {\bibinfo  {journal} {Phys. Fluids}\ }\textbf {\bibinfo
  {volume} {11}},\ \bibinfo {pages} {943} (\bibinfo {year} {1999})}\BibitemShut
  {NoStop}%
\bibitem [{\citenamefont {Kaneda}\ and\ \citenamefont
  {Yamamoto}(2021)}]{kaneda2021}%
  \BibitemOpen
  \bibfield  {author} {\bibinfo {author} {\bibfnamefont {Y.}~\bibnamefont
  {Kaneda}}\ and\ \bibinfo {author} {\bibfnamefont {Y.}~\bibnamefont
  {Yamamoto}},\ }\bibfield  {title} {\bibinfo {title} {Velocity {{Gradient
  Statistics}} in {{Turbulent Shear Flow}}: {{An Extension}} of
  {{Kolmogorov}}'s {{Local Equilibrium Theory}}},\ }\href@noop {} {\bibfield
  {journal} {\bibinfo  {journal} {J. Fluid Mech.}\ }\textbf {\bibinfo {volume}
  {929}} (\bibinfo {year} {2021})}\BibitemShut {NoStop}%
\bibitem [{\citenamefont {Hoyas}\ \emph {et~al.}(2022)\citenamefont {Hoyas},
  \citenamefont {Oberlack}, \citenamefont {Alcántara-Ávila}, \citenamefont
  {Kraheberger},\ and\ \citenamefont {Laux}}]{hoyas2022}%
  \BibitemOpen
  \bibfield  {author} {\bibinfo {author} {\bibfnamefont {S.}~\bibnamefont
  {Hoyas}}, \bibinfo {author} {\bibfnamefont {M.}~\bibnamefont {Oberlack}},
  \bibinfo {author} {\bibfnamefont {F.}~\bibnamefont {Alcántara-Ávila}},
  \bibinfo {author} {\bibfnamefont {S.~V.}\ \bibnamefont {Kraheberger}},\ and\
  \bibinfo {author} {\bibfnamefont {J.}~\bibnamefont {Laux}},\ }\bibfield
  {title} {\bibinfo {title} {Wall {{Turbulence}} at {{High Friction Reynolds
  Numbers}}},\ }\href@noop {} {\bibfield  {journal} {\bibinfo  {journal} {Phys.
  Rev. Fluids}\ }\textbf {\bibinfo {volume} {7}},\ \bibinfo {pages} {014602}
  (\bibinfo {year} {2022})}\BibitemShut {NoStop}%
\bibitem [{\citenamefont {Zhang}\ \emph {et~al.}(2018)\citenamefont {Zhang},
  \citenamefont {Song}, \citenamefont {Ye}, \citenamefont {Wang}, \citenamefont
  {Huang}, \citenamefont {An},\ and\ \citenamefont {Chen}}]{zhang2018}%
  \BibitemOpen
  \bibfield  {author} {\bibinfo {author} {\bibfnamefont {Z.}~\bibnamefont
  {Zhang}}, \bibinfo {author} {\bibfnamefont {X.}~\bibnamefont {Song}},
  \bibinfo {author} {\bibfnamefont {S.}~\bibnamefont {Ye}}, \bibinfo {author}
  {\bibfnamefont {Y.}~\bibnamefont {Wang}}, \bibinfo {author} {\bibfnamefont
  {C.}~\bibnamefont {Huang}}, \bibinfo {author} {\bibfnamefont
  {Y.}~\bibnamefont {An}},\ and\ \bibinfo {author} {\bibfnamefont
  {Y.}~\bibnamefont {Chen}},\ }\bibfield  {title} {\bibinfo {title}
  {Application of {{Deep Learning Method}} to {{Reynolds Stress Models}} of
  {{Channel Flow Based}} on {{Reduced-Order Modeling}} of {{{DNS} Data}}},\
  }\href@noop {} {\bibfield  {journal} {\bibinfo  {journal} {J. Hydrodyn.}\
  }\textbf {\bibinfo {volume} {31}} (\bibinfo {year} {2018})}\BibitemShut
  {NoStop}%
\bibitem [{\citenamefont {Goodfellow}\ \emph {et~al.}(2016)\citenamefont
  {Goodfellow}, \citenamefont {Bengio},\ and\ \citenamefont
  {Courville}}]{Goodfellow-et-al-2016}%
  \BibitemOpen
  \bibfield  {author} {\bibinfo {author} {\bibfnamefont {I.}~\bibnamefont
  {Goodfellow}}, \bibinfo {author} {\bibfnamefont {Y.}~\bibnamefont {Bengio}},\
  and\ \bibinfo {author} {\bibfnamefont {A.}~\bibnamefont {Courville}},\
  }\href@noop {} {\emph {\bibinfo {title} {Deep Learning}}}\ (\bibinfo
  {publisher} {{MIT Press}},\ \bibinfo {year} {2016})\BibitemShut {NoStop}%
\bibitem [{\citenamefont {Angeli}\ \emph {et~al.}(2015)\citenamefont {Angeli},
  \citenamefont {Bieder},\ and\ \citenamefont {Fauchet}}]{angeli2015}%
  \BibitemOpen
  \bibfield  {author} {\bibinfo {author} {\bibfnamefont {P.-E.}\ \bibnamefont
  {Angeli}}, \bibinfo {author} {\bibfnamefont {U.}~\bibnamefont {Bieder}},\
  and\ \bibinfo {author} {\bibfnamefont {G.}~\bibnamefont {Fauchet}},\
  }\bibfield  {title} {\bibinfo {title} {Overview of the {{TrioCFD Code}}:
  {{Main Features}}, {{V}}\&v {{Procedures}} and {{Typical Applications}} to
  {{Nuclear Engineering}}},\ }in\ \href@noop {} {\emph {\bibinfo {booktitle}
  {Proceedings of 16th {{International Topical Meeting}} on {{Nuclear Reactor
  Thermal Hydraulics}} ({{NURETH-16}})}}}\ (\bibinfo {year} {2015})\BibitemShut
  {NoStop}%
\bibitem [{\citenamefont {Angeli}\ \emph {et~al.}(2017)\citenamefont {Angeli},
  \citenamefont {Puscas}, \citenamefont {Fauchet},\ and\ \citenamefont
  {Cartalade}}]{angeli2017}%
  \BibitemOpen
  \bibfield  {author} {\bibinfo {author} {\bibfnamefont {P.-E.}\ \bibnamefont
  {Angeli}}, \bibinfo {author} {\bibfnamefont {M.-A.}\ \bibnamefont {Puscas}},
  \bibinfo {author} {\bibfnamefont {G.}~\bibnamefont {Fauchet}},\ and\ \bibinfo
  {author} {\bibfnamefont {A.}~\bibnamefont {Cartalade}},\ }\bibfield  {title}
  {\bibinfo {title} {{FVCA8} {{Benchmark}} for the {{Stokes}} and
  {{Navier-Stokes Equations}} with the {{TrioCFD Code}} – {{Benchmark
  Session}}},\ }in\ \href@noop {} {\emph {\bibinfo {booktitle} {Finite
  {{Volumes}} for {{Complex Applications}} 8}}}\ (\bibinfo {address} {{Lille,
  France}},\ \bibinfo {year} {2017})\BibitemShut {NoStop}%
\end{thebibliography}%

\end{document}